\def\tr{{\rm tr}\,}
\def\Tr{{\rm Tr}\,}
\def\wt{\widetilde}

\def\b{\bibitem}
\documentstyle[aps,prb,eqsecnum,psfig,floats]{revtex}
\begin{document}
\def\SNG{{\em Physical Review Style and Notation Guide}}
\def\LUG {{\em \LaTeX{} User's Guide \& Reference Manual}}
\def\btt#1{{\tt$\backslash$\string#1}}%
\def\REVTeX{REV\TeX}
\def\AmS{{\protect\the\textfont2
        A\kern-.1667em\lower.5ex\hbox{M}\kern-.125emS}}
\def\AmSLaTeX{\AmS-\LaTeX}
\def\BibTeX{\rm B{\sc ib}\TeX}
\twocolumn[\hsize\textwidth\columnwidth\hsize\csname@twocolumnfalse%
\endcsname
 
\title{Properties of spin-triplet, even-parity superconductors\\
}
\author{D.Belitz}
\address{Department of Physics and Materials Science Institute\\
University of Oregon,\\
Eugene, OR 97403}
\author{T.R.Kirkpatrick}
\address{Institute for Physical Science and Technology, and Department of 
 Physics\\
 University of Maryland,\\ 
 College Park, MD 20742}

\date{\today}
\maketitle

\begin{abstract}
The physical consequences of the spin-triplet, even-parity pairing that has
been predicted to exist in disordered two-dimensional electron systems are
considered in detail. We show that the presence of an attractive interaction
in the particle-particle spin-triplet channel leads to an instability of
the normal metal that competes with the localizing effects of the disorder.
The instability is characterized by a diverging length scale, and has all
of the characteristics of a continuous phase transition. The transition
and the properties of the ordered phase are studied in mean-field theory,
and by taking into account Gaussian fluctuations. We find that the ordered
phase is indeed a superconductor with an ordinary Meissner effect and a
free energy that is lower than that of the normal metal. Various
technical points that have given rise to confusion in connection with
this and other manifestations of odd-gap superconductivity are also discussed.
\end{abstract}
]

\section{Introduction}
\label{sec:I}

The possibility of superconductivity with a gap function that is an odd
function of time or frequency has been the subject of some attention
lately. The concept was introduced by Berezinskii\cite{Berezinskii} in
the context of $^3$He, and it was more recently revived in connection
with two-dimensional ($2$-$D$) electron fluids in 
semiconductors,\cite{us_mosfets}
and with high-$T_{\rm c}$ superconductors.\cite{Abrahamsetal} In 
Berezinskii's original work, as well as in 
Refs.\ \onlinecite{us_mosfets,us_tsc},
the case of spin-triplet, even-parity pairing was considered, while
Refs.\ \onlinecite{Abrahamsetal} also discussed the case of spin-singlet,
odd-parity pairing. In either case, the gap function being an odd function
of the frequency ensures that the Pauli principle is obeyed.

In these references, some properties of odd-gap superconductors have been
explored, but no complete analysis of the phase transition, or of the
superconducting phase, has been given. Furthermore, several fundamental
questions concerning the stability of such a superconducting phase,
and whether an odd-gap superconductor can indeed be superconducting,
have led to considerable confusion.

In the present paper we discuss these issues. For definiteness, we will
analyze the mechanism for spin-triplet, even-parity superconductivity
discovered in Refs.\ \onlinecite{us_tsc}, but many of our conclusions
apply to other odd-gap superconductors as well. The paper is organized as
follows. In Sec.\ \ref{sec:II} we give a field-theoretic formulation of
the problem and discuss the stability properties of the field theory.
In Sec.\ \ref{sec:III} we show that in the presence of an attractive
interaction in the particle-particle spin-triplet channel, there is an
instability of the normal conducting phase. This instability has all
of the properties of a second order phase transition, with a diverging
length scale, a diverging order parameter susceptibility, etc.
In Sec.\ \ref{sec:IV} we develop and analyze a mean-field theory of the
transition. In particular, we determine the critical behavior on either
side of the transition. We then determine the properties
of the ordered phase in a Gaussian approximation. We show that the ordered
phase is a superconductor, with a Meissner effect, and a real part of the
conductivity that has a delta-function contribution, just as in BCS
superconductors. In Sec.\ \ref{sec:V}
we conclude with a summary and a general discussion of our results.

\section{Field-theoretic formulation of the problem}
\label{sec:II}

\subsection{$Q$-matrix theory for fermions}
\label{subsec:II.A}

Our starting point is a general field theory for electrons. For any
fermionic system, the partition function can be written\cite{NegeleOrland}
\begin{mathletters}
\label{eqs:2.1}
\begin{equation}
Z = \int D[{\bar\psi},\psi]\ \exp\left(S[{\bar\psi},
                       \psi] \right)\quad,
\label{eq:2.1a}
\end{equation}
where $S$ is the action in terms of the fermionic (i.e., Grassmann valued)
fields $\bar\psi$ and $\psi$. We consider an action that consists of a
free-fermion part $S_0$, a part $S_{\rm dis}$ describing the interaction
of the electrons with quenched disorder, and a part $S_{\rm int}$ describing
the electron-electron interaction,
\begin{equation}
S = S_0 + S_{\rm dis} + S_{\rm int}\quad,
\label{eq:2.1b}
\end{equation}
\end{mathletters}%
Each field $\psi$ or $\bar\psi$ carries a Matsubara frequency
index $n$, a spin index $\sigma=\uparrow,\downarrow$, and, if the quenched
disorder is dealt with by means of the replica trick,
a replica index $\alpha$. For our purposes it is useful to 
introduce a matrix of bilinear products of the fermion fields,
\begin{eqnarray}
B_{12} &=& \frac{i}{2}\,\left( \begin{array}{cccc}
          -\psi_{1\uparrow}{\bar\psi}_{2\uparrow} &
             -\psi_{1\uparrow}{\bar\psi}_{2\downarrow} &
                 -\psi_{1\uparrow}\psi_{2\downarrow} &
                      \ \ \psi_{1\uparrow}\psi_{2\uparrow}  \\
          -\psi_{1\downarrow}{\bar\psi}_{2\uparrow} &
             -\psi_{1\downarrow}{\bar\psi}_{2\downarrow} &
                 -\psi_{1\downarrow}\psi_{2\downarrow} &
                      \ \ \psi_{1\downarrow}\psi_{2\uparrow}  \\
          \ \ {\bar\psi}_{1\downarrow}{\bar\psi}_{2\uparrow} &
             \ \ {\bar\psi}_{1\downarrow}{\bar\psi}_{2\downarrow} &
                 \ \ {\bar\psi}_{1\downarrow}\psi_{2\downarrow} &
                      -{\bar\psi}_{1\downarrow}\psi_{2\uparrow} \\
          -{\bar\psi}_{1\uparrow}{\bar\psi}_{2\uparrow} &
             -{\bar\psi}_{1\uparrow}{\bar\psi}_{2\downarrow} &
                 -{\bar\psi}_{1\uparrow}\psi_{2\downarrow} &
                      \ \ {\bar\psi}_{1\uparrow}\psi_{2\uparrow} \\
                    \end{array}\right)
\nonumber\\
&\cong& Q_{12}\quad.
\label{eq:2.2}
\end{eqnarray}
where all fields are understood to be taken at position ${\bf x}$, and
$1\equiv (n_1,\alpha_1)$, etc. The
matrix elements of $B$ commute with one another, and are therefore
isomorphic to classical or number-valued fields that we denote by 
$Q$.\cite{Notation_Footnote}
This isomorphism maps the adjoint operation on products of fermion fields,
which is denoted above by an overbar, on the complex conjugation of the
classical fields. We use the isomorphism to
constrain $B$ to the classical field $Q$, and exactly rewrite the partition
function\cite{us_fermions}
\begin{eqnarray}
Z &=& \int D[{\bar\psi},\psi]\ e^{S[{\bar\psi},\psi]}
      \int D[Q]\,\delta[Q-B]
\nonumber\\
  &=& \int D[{\bar\psi},\psi]\ e^{S[{\bar\psi},\psi]}
      \int D[Q]\,D[{\wt\Lambda}]\ e^{\Tr [{\wt\Lambda}(Q-B)]}
\nonumber\\
  &\equiv& \int D[Q]\,D[{\wt\Lambda}]\ e^{{\cal A}[Q,{\wt\Lambda}]}\quad.
\label{eq:2.3}
\end{eqnarray}
${\wt\Lambda}$ is an auxiliary bosonic matrix field that serves to enforce the
functional delta-constraint in the first line of Eq.\ (\ref{eq:2.3}),
and the last line defines the action ${\cal A}$.
The matrix elements of both $Q$ and ${\wt\Lambda}$
are spin-quaternions (i.e., elements of ${\cal Q}\times{\cal Q}$ with
${\cal Q}$ the quaternion field). From Eq.\ (\ref{eq:2.2}) we see that
expectation values of the $Q$ matrix elements yield 
local Green functions, and $Q$-$Q$
correlation functions describe four-fermion correlation functions.
The physical meaning of ${\wt\Lambda}$ 
is that its expectation value plays the role of a
self energy (see Ref.\ \onlinecite{us_fermions} and Sec.\ \ref{sec:IV}
below). 

It is convenient to expand the $4\times 4$ matrix in Eq.\ (\ref{eq:2.2})
in a spin-quaternion basis,
\begin{equation}
Q_{12}({\bf x}) = \sum_{r,i=0,3} (\tau_r\otimes s_i)\,{^i_rQ_{12}}({\bf x})
                 \quad
\label{eq:2.4}
\end{equation}
and analogously for $\wt\Lambda$. Here 
$\tau_0 = s_0 = \openone_2$ is the
$2\times 2$ unit matrix, and $\tau_j = -s_j = -i\sigma_j$, $(j=1,2,3)$,
with $\sigma_{1,2,3}$ the Pauli matrices. In this basis, $i=0$ and $i=1,2,3$
describe the spin singlet and the spin triplet, respectively. An explicit
calculation reveals that $r=0,3$ corresponds to the particle-hole channel
(i.e., products ${\bar\psi}\psi$), while $r=1,2$ describes the
particle-particle channel (i.e., products ${\bar\psi}{\bar\psi}$ or
$\psi\psi$). We will be particularly interested in the matrix elements of
${^1_1 Q}$, for which the isomorphism expressed in Eq.\ (\ref{eq:2.2}) 
reads\cite{OP_Footnote}
\begin{eqnarray}
{^1_1 Q}_{12}({\bf x})&\cong&\frac{i}{8}\,\left[
          \psi_{1\uparrow}({\bf x})\psi_{2\uparrow}({\bf x})
   - \psi_{1\downarrow}({\bf x})\psi_{2\downarrow}({\bf x})\right.
\nonumber\\
&& + \left.{\bar\psi}_{1\downarrow}({\bf x}){\bar\psi}_{2\downarrow}({\bf x})
   - {\bar\psi}_{1\uparrow}({\bf x}){\bar\psi}_{2\uparrow}({\bf x})\right]
                                                     \quad.
\label{eq:2.4'}
\end{eqnarray}
From the structure of Eq.\ (\ref{eq:2.2}) one obtains the
following formal symmetry properties of the $Q$ matrices,\cite{us_fermions}
\begin{mathletters}
\label{eqs:2.5}
\begin{eqnarray}
{^0_r Q}_{12}&=&(-)^r\,{^0_r Q}_{21}\quad,\quad (r=0,3)\quad,
\label{eq:2.5a}\\
{^i_r Q}_{12}&=&(-)^{r+1}\,{^i_r Q}_{21}\ ,\ (r=0,3;\ i=1,2,3)\quad,
\label{eq:2.5b}\\
{^0_r Q}_{12}&=&{^0_r Q}_{21}\quad,\quad (r=1,2)\quad,
\label{eq:2.5c}\\
{^i_r Q}_{12}&=&-{^i_r Q}_{21}\quad,\quad (r=1,2;\ i=1,2,3)\quad,
\label{eq:2.5d}\\
{^i_r Q}_{12}^*&=&- {^i_r Q}_{-n_1-1,-n_2-1}^{\alpha_1\alpha_2}\quad.
\label{eq:2.5e}
\end{eqnarray}
\end{mathletters}%
Here the star in Eq.\ (\ref{eq:2.5e}) denotes complex conjugation.

By using the delta constraint in Eq.\ (\ref{eq:2.3}) to rewrite all terms 
that are quartic in the fermion field in terms of $Q$, we can achieve
an integrand that is bilinear in $\psi$ and $\bar\psi$. The Grassmannian
integral can then be performed exactly, and we obtain for the
action ${\cal A}$
\begin{mathletters}
\label{eqs:2.6}
\begin{eqnarray}
{\cal A}[Q,{\wt\Lambda}] &=& {\cal A}_{\rm dis} + {\cal A}_{\rm int}
                           + \frac{1}{2}\,\Tr\ln\left(G_0^{-1} - i{\wt\Lambda}
                                       \right)
\nonumber\\
  && + \int d{\bf x}\ \tr\left({\wt\Lambda}({\bf x})\,Q({\bf x})\right)\quad.
\label{eq:2.6a}
\end{eqnarray}
Here
\begin{equation}
G_0^{-1} = -\partial_{\tau} + \partial_{\bf x}^2/2m + \mu\quad,
\label{eq:2.6b}
\end{equation}
\end{mathletters}%
is the inverse free electron Green operator, with $\partial_{\tau}$ and
$\partial_{\bf x}$ derivatives with respect to imaginary time and position,
respectively, $m$ is the electron mass, and $\mu$ is the chemical potential.
$\Tr$ denotes a trace over all degrees of freedom, including the continuous
position variable, while $\tr$ is a trace over all those discrete indices that
are not explicitly shown. The electron-electron interaction 
${\cal A}_{\rm int}$ is conveniently decomposed into four pieces 
that describe the interaction
in the particle-hole and particle-particle spin-singlet and spin-triplet 
channels.\cite{us_fermions} 
For the purposes of the present paper, we need only the
particle-particle spin-triplet channel interaction explicitly. In 
Ref.\ \onlinecite{us_tsc} it was shown that in any quenched disordered,
interacting electron system, there is an attractive interaction
in the particle-particle spin-triplet channel of the form
\begin{mathletters}
\label{eqs:2.7}
\begin{eqnarray}
{\cal A}_{\rm int}^{\rm p-p,t} = -\pi N_{\rm F}\int d{\bf x}\ T\sum_{n_1,n_2,
   n_3, n_4} \delta_{n_1+n_2,n_3+n_4}
\nonumber\\
\times {\wt K}_{n_1,n_2;n_3,n_4} \sum_{r=1,2} \sum_{i=1}^{3}\sum_{\alpha} 
   {^i_rQ}_{n_1n_2}^{\alpha\alpha}\,{^i_rQ}_{n_3n_4}^{\alpha\alpha}\quad.
\label{eq:2.7a}
\end{eqnarray}
In $D=2$, the effective interaction potential is
\begin{eqnarray}
{\wt K}_{n_1,n_2;n_3,n_4}&=&\frac{1}{4}\,\left(K_{n_1,n_2;n_3,n_4}
   - K_{n_2,n_1;n_3,n_4} \right.
\nonumber\\
 &&\left. - K_{n_1,n_2;n_4,n_3} + K_{n_2,n_1;n_4,n_3}\right)\quad,
\label{eq:2.7b}
\end{eqnarray}
where\cite{PotentialFootnote}
\begin{equation}
K_{n_1,n_2;n_3,n_4} = y\,\ln\left\vert\frac{n_1-n_3}{n_2-n_3}\right\vert
   \quad,
\label{eq:2.7c}
\end{equation}
\end{mathletters}%
with a positive coupling constant $y>0$ that depends both on the
disorder strength and on the coupling constants in the 
particle-hole interaction channels. Although both the disorder and 
the particle-hole
channel interactions are necessary to produce the particle-particle
spin-triplet interaction, none of the points to be investigated in this
paper qualitatively depends on either one of them other than through the
existence of $\wt K$. For the Gaussian theory that we will consider, the 
only other effect of the disorder that is
relevant to our dicussion is that it replaces some free-electron
correlation functions by diffusive ones, which changes the exponents
in certain scaling relations.
For notational simplicity, and in order to keep 
our discussion technically as simple as
possible, in what follows we therefore neglect all contributions
to the action that are nonessential for our purposes. We thus
work with a system given by Eq.\ (\ref{eq:2.6a}) with
${\cal A}_{\rm dis} = 0$ and ${\cal A}_{\rm int} = {\cal A}_{\rm int}^{p-p,t}$
and drop the replica indices on all fields. In cases where diffusive
correlations make a difference, we will mention this explicitly and restore 
diffusive scaling. Our restriction to a Gaussian approximation
purposely neglects the localizing effects of the 
disorder. We will come back to this point in Sec.\ \ref{subsec:V.C}.

Note that Eq. (\ref{eq:2.7c}) means that in time space, the interaction
between superconducting fluctuations is long-ranged.
Consequently, the critical behavior, and in particular the critical exponents
discussed in Secs.\ \ref{sec:III} and \ref{subsec:IV.A} below, 
depend explicity on the detailed form of the kernel.
However, our qualitative results, in particular the spontaneous symmetry
breaking and the existence of an
ordinary quantum critical point that marks the onset of superconducting
long-range order, we expect to be generic. We also note that in
Eqs.\ (\ref{eqs:2.7}) above, as in Ref.\ \onlinecite{us_tsc}, we have
neglected any wavenumber dependences of the effective interaction potential.
Any nontrivial, i.e. nonanalytic, wavenumber dependence would correspond to
an interaction that is long-ranged in real space. Such long-ranged interactions
are known to stabilize mean-field critical behavior,\cite{FisherMaNickel}
and will thus increase the range of validity of our mean-field theory.
We therefore do not expect our neglecting the wavenumber dependence of $K$ to
qualitatively affect our results.

\subsection{The Fermi-liquid saddle point, and Gaussian fluctuations}
\label{subsec:II.B}

In Ref.\ \onlinecite{us_fermions} it was shown that the above 
$Q$-${\wt\Lambda}$ field theory possesses a saddle-point solution that
describes a free Fermi gas (or a disordered Fermi liquid if we had not
dropped the disorder and particle-hole channel interaction contributions
to the action). That is, the saddle-point equations
\begin{equation}
\frac{\delta {\cal A}}{\delta Q}\bigg\vert_{Q_{\rm sp},\wt\Lambda_{\rm sp}}
= \frac{\delta {\cal A}}{\delta\wt\Lambda}\bigg\vert_{Q_{\rm sp},
                      \wt\Lambda_{\rm sp}} = 0\quad,
\label{eq:2.8}
\end{equation}
are solved by the {\it ansatz}
\begin{mathletters}
\label{eqs:2.9}
\begin{eqnarray}
{_r^iQ}_{12}({\bf x})\Bigl\vert_{\rm sp}&=&\delta_{12}
   \,\delta_{r0}\,\delta_{i0}\,Q_{n_1}\quad,
\label{eq:2.9a}\\
{_r^i\wt\Lambda}_{12}({\bf x})\Bigl\vert_{\rm sp}&=&\delta_{12}\,\delta_{r0}\,
                                           \delta_{i0}\,\Lambda_{n_1}\quad,
\label{eq:2.9b}
\end{eqnarray}
\end{mathletters}%
with
\begin{mathletters}
\label{eqs:2.10}
\begin{equation}
Q_n=\frac{i}{2V}\,\sum_{\bf p}\,G_n^0({\bf p})\quad,
\label{eq:2.10a}
\end{equation}
\begin{equation}
\Lambda_n = 0\quad.
\label{eq:2.10b}
\end{equation}
\end{mathletters}%
The saddle-point Green function $G_n^0$ is simply the one for free electrons,
\begin{equation}
G_n^0({\bf p}) = \left(i\omega_n - \xi_{\bf p} \right)^{-1} \quad,
\label{eq:2.11}
\end{equation}
with $\xi_{\bf p} = {\bf p}^2/2m - \mu$.

We next consider the Gaussian fluctuations about this saddle point.
Proceeding as in Ref.\ \onlinecite{us_fermions}, we write
\begin{mathletters}
\label{eqs:2.12}
\begin{eqnarray}
Q = Q_{\rm sp} + \delta Q\quad,
\label{eq:2.12a}\\
\tilde\Lambda = \tilde\Lambda_{\rm sp} + \delta\tilde\Lambda\quad,
\label{eq:2.12b}
\end{eqnarray}
\end{mathletters}%
and introduce a new field ${\bar\Lambda}$ by
\begin{mathletters}
\label{eqs:2.13}
\begin{equation}
\bar\Lambda_{12} = \frac{1}{2}\,\varphi_{12}\,{\wt\Lambda}_{12}
                   + Q_{12}\quad,
\label{eq:2.13a}
\end{equation}
with
\begin{equation}
\varphi_{nm}({\bf k}) = \frac{1}{V}\sum_{\bf p}\,
  G_{\rm sp}({\bf p},\omega_{n})\,G_{\rm sp}({\bf p}+{\bf k,}\omega_{m})\quad,
\label{eq:2.13b}
\end{equation}
\end{mathletters}%
a generalized Lindhard function.
$\bar\Lambda$ has been chosen so that $Q$ and $\bar\Lambda$ decouple in
the Gaussian action. Expanding to second order in the fluctuating fields
$\delta Q$ and $\delta{\bar\Lambda}$, we obtain the Gaussian action
\begin{eqnarray}
{\cal A}_{\rm G} &=& \frac{4}{V}\sum_{\bf k} \sum_{1,2}\sum_{r,i}
   \left({{+\atop -}\atop{-\atop +}}\right)_r\,\varphi_{12}^{-1}({\bf k})
   \left[\,{^i_r(}\delta{\bar\Lambda})_{12}({\bf k})\,\right.
\nonumber\\
&&\left. \times\,{^i_r(}\delta{\bar\Lambda})_{12}(-{\bf k})
    -\,{^i_r(}\delta Q)_{12}({\bf k})\,
 ^i_r(\delta Q)_{12}(-{\bf k})\right]
\nonumber\\
 &&\qquad\qquad\qquad\qquad\qquad + {\cal A}_{\rm int}^{\rm p-p,t}[\delta Q] 
                                                       \quad,
\label{eq:2.14}
\end{eqnarray}
where the symbol $\left({{+\atop -}\atop{-\atop +}}\right)_r$
is equal to $+1$ for $r=0,3$, and $-1$ for $r=1,2$.

Keeping the disorder part of the action results in the inverse Lindhard
function $\varphi_{12}^{-1}({\bf k})$ in Eq.\ (\ref{eq:2.14}) being replaced
by\cite{us_fermions}
\begin{mathletters}
\label{eqs:2.15}
\begin{equation}
{\cal D}^{-1}_{12}({\bf k}) = \varphi^{-1}_{12}({\bf k}) - 1/\pi\,N_{\rm F}\,
   \tau\quad,
\label{eq:2.15a}
\end{equation}
with $\tau$ the elastic scattering mean-free time. For small wavenumbers
and small $\omega_{n_1}- \omega_{n_2}$ with $n_1n_2<0$, ${\cal D}$ is
diffusive,
\begin{equation}
{\cal D}_{12}({\bf k}) = \frac{\pi\,N_{\rm F}}{D\,{\bf k}^2 + \vert
   \omega_{n_1} - \omega_{n_2}\vert}\quad,
\label{eq:2.15b}
\end{equation}
\end{mathletters}%
with $D$ the Boltzmann value of the diffusion constant. Whenever it makes
a difference in the results, we will use the diffusive propagator
${\cal D}$ instead of the free fermion propagator $\varphi$.

In order to completely define the field theory, one also must specify the
integration contours for the functional integral, Eq.\ (\ref{eq:2.3}).
The most obvious choice would be to integrate over the space of
all matrices that obey Eqs.\ (\ref{eqs:2.5}). However, this creates the
following problem. According to Eqs.\ (\ref{eq:2.5d}), (\ref{eq:2.5e}),
${^1_1Q}_{n-1,-n}$ is real, and so are all other particle-particle spin-triplet
components of $Q_{n-1,-n}$. However, according to Eq.\ (\ref{eq:2.14}),
this direction in the complex $Q$-plane is unstable, and
the direction of steepest descent is along the imaginary
axis.\cite{Imaginary_Footnote}
In order to do the integral by the saddle-point method,
the contour for the integration over
the ${^{1,2,3}_{1,2}Q}$ must therefore be
deformed so that it passes through the saddle point in the direction of the
imaginary axis. At least within perturbation theory, this just amounts
to formally doing the Gaussian integral without worrying about
convergence problems. A necessary condition for this procedure to be valid
is that the resulting theory reproduces the perturbative results obtained
within the underlying fermionic
theory. An easy and convenient check is provided by the so-called
weak-localization correction to the conductivity of non-interacting
electrons. For our purposes, a simple structural check suffices and is
provided in Appendix\ \ref{app:A}. The check is affirmative, and suggests
that we can safely ignore convergence questions relating to the
Gaussian integrals that occur in our field theory.
A related point will become important in Sec.\ \ref{sec:IV} below.

\section{Instability of the normal metal}
\label{sec:III}

\subsection{The Gaussian eigenvalue problem in the normal phase}
\label{subsec:III.A}

We now turn to the full Gaussian action in an expansion about the Fermi 
liquid saddle-point, Eq.\ (\ref{eq:2.14}). Integrating out the
auxiliary field ${\bar\Lambda}$ just contributes a multiplicative constant
to the partition function, and we are left with a quadratic form in
$\delta Q$. Let us consider the particle-particle spin-triplet part of the 
action, e.g., the channel $r=i=1$. Since the action is diagonal in all
indices except the frequency, we can drop the $r$ and $i$ indices as
well as the wavenumber dependence, and consider a quadratic form
\begin{mathletters}
\label{eqs:3.1}
\begin{equation}
\sum_{1,2,3,4} \delta Q_{12}\,M_{12,34}\,\delta Q_{34}\quad,
\label{eq:3.1a}
\end{equation}
where the matrix $M$ has the structure
\begin{equation}
M_{12,34} = -a_{12}\,\delta_{13}\,\delta_{24} 
       - b_{1-3,2-3}\,\delta_{1+2,3+4} \quad.
\label{eq:3.1b}
\end{equation}
\end{mathletters}%
Here $b$ is proportional to the interaction potential $K$, 
Eq.\ (\ref{eq:2.7c}), and $a_{12}$ is either $\varphi^{-1}_{12}$,
Eq.\ (\ref{eq:2.13b}), or ${\cal D}^{-1}_{12}$, Eq.\ (\ref{eq:2.15a}),
depending on whether or not we keep the disorder explicitly.
It is useful to consider $M$ as a matrix with
composite indices $(1,2)$ and $(3,4)$, and to study the eigenvalue
problem
\begin{equation}
\sum_{1',2'} M_{12,1'2'}\,f_{1'2'} = \lambda\,f_{12}\quad,
\label{eq:3.2}
\end{equation}
with eigenvalues $\lambda$ and eigenfunctions $f_{12}$. For reasons
that will soon become apparent, it is advantageous to transform to
a different basis of eigenfunctions $g$ defined by
\begin{equation}
g_{12} = \sum_{3,4} b_{1-3,2-3}\,\delta_{1+2,3+4}\,f_{34}\quad.
\label{eq:3.3}
\end{equation}
In terms of the $g$, the eigenvalue problem reads
\begin{mathletters}
\label{eqs:3.4}
\begin{equation}
g_{12} = -\sum_{1'}\frac{b_{1-1',2-1'}}{a_{1',-1'+(1+2)} + \lambda}\ 
         g_{1',-1'+(1+2)}\quad.
\label{eq:3.4a}
\end{equation}
Now we write $\omega_{n_2} = -\omega_{n_1} + \Omega_n$, analytically
continue to real frequencies at $T=0$, and put the wavenumber dependence
back in. Then we obtain the eigenvalue equation in the form
\begin{eqnarray}
g_{\lambda}(\omega;{\bf p},\Omega) &=& -y\int_{-\infty}^{\infty} dx\ 
   \ln\left\vert\frac{\omega - x}{-\omega - x + \Omega}\right\vert
\nonumber\\
&&\quad\times \frac{1}{a(x;{\bf p},\Omega) + \lambda}\ 
     g_{\lambda}(x;{\bf p},\Omega)\quad.
\label{eq:3.4b}
\end{eqnarray}
\end{mathletters}%
Here $y$ is the coupling constant of Eq.\ (\ref{eq:2.7c}), scaled by
an appropriate factor.
In order to check for an instability of the metallic phase, we need to
look for a zero eigenvalue. From the structure of Eq.\ (\ref{eq:3.4b})
it is clear that the first zero eigenvalue appears for ${\bf p}=\Omega=0$.
We therefore specialize to zero momentum and external frequency. We then
have $a(x;{\bf p}=0,\Omega=0) \propto \vert x\vert$, irrespective of
whether we use $a=\varphi^{-1}$ or $a={\cal D}^{-1}$. Absorbing a
constant factor into the eigenvalue $\lambda$ and into the coupling
constant $y$, we then get the eigenvalue equation in the form
\begin{equation}
g_{\lambda}(\omega) = y \int_{0}^{\infty} \frac{dx}{x + \lambda}\,
   \ln\left\vert\frac{x + \omega}{x - \omega}\right\vert\,g_{\lambda}(x)\quad,
\label{eq:3.5}
\end{equation}
with $g_{\lambda}(\omega) \equiv g_{\lambda}(\omega;{\bf p}=0,\Omega=0)$.
Notice that Eq.\ (\ref{eq:3.5}) is a generalization of the gap equation
in Ref.\ \onlinecite{us_tsc}: The critical eigenfunction, 
$g_{\lambda=0}(\omega)$, obeys the same integral equation as the 
critical gap function. This is the first indication that the long-range
order implied by a nonzero gap function in Ref.\ \onlinecite{us_tsc}
can indeed be understood in terms of a conventional continuous phase
transition that is triggered by an instability of the metallic phase. 

\subsection{Solution of the eigenvalue problem}
\label{subsec:III.B}

The integral equation, Eq.\ (\ref{eq:3.5}), is very hard to solve analytically.
We therefore make the same approximation as was done for the gap equation
in Ref.\ \onlinecite{us_tsc}, namely replacing the logarithmic kernel
by a rational one with a similar overall behavior:
\begin{equation}
\ln\left\vert\frac{x + \omega}{x - \omega}\right\vert \rightarrow
   \frac{2x}{\omega}\,\Theta (x-\omega) + \frac{2\omega}{x}\,\Theta(\omega-x)
   \quad.
\label{eq:3.6}
\end{equation}
This allows to rewrite the integral equation as an ordinary differential
equation,
\begin{equation}
\frac{d^2}{d\omega^2}\,g_{\lambda}(\omega)
  + \frac{1}{\omega}\,\frac{d}{d\omega}\,g_{\lambda}(\omega)
  + \left[\frac{4y}{\omega(\omega + \lambda)} - \frac{1}{\omega^2}\right]
                       g_{\lambda}(\omega) = 0.
\label{eq:3.6'}
\end{equation}
It is useful to rewrite this ODE in the form
\begin{mathletters}
\label{eqs:3.7}
\begin{equation}
\frac{d^2}{dz^2}\,w(z) + p(z)\,\frac{d}{dz}\,w(z) + q(z)\,w(z) = 0\quad,
\label{eq:3.7a}
\end{equation}
with
\begin{equation}
p(z) = \frac{1}{z}\quad,
\label{eq:3.7b}
\end{equation}
\begin{equation}
q(z) = \frac{-1}{z^2} + \frac{t-1}{z} + \frac{1-t}{z-1}\quad,
\label{eq:3.7c}
\end{equation}
where $t=1-4y$, and the solution $w(z)$ of the ODE determines the eigenfunction
via
\begin{equation}
g_{\lambda}(\omega) = \lambda^2\,w(-\omega/\lambda)\quad.
\label{eq:3.7d}
\end{equation}
\end{mathletters}%

The ODE, Eq.\ (\ref{eq:3.7a}), is Fuchsian with three regular singular points
at $z=0$, $z=1$, and $z=\infty$. It can thus be transformed into a
hypergeometric equation.\cite{DenneryKrzywicki} Before we look at the
general solution, let us consider the case $\lambda=0$, where the two
solutions of Eq.\ (\ref{eq:3.6'}) are
\begin{equation}
g_{\lambda=0}^{(1)} = \omega^{-\sqrt t}\quad,\quad
   g_{\lambda=0}^{(2)} = \omega^{\sqrt t}\quad.
\label{eq:3.8}
\end{equation}
In order to obtain a solution that is well-behaved everywhere, we therefore
must require $t(\lambda=0)=0$. More generally, the requirement of a
well-behaved eigenfunction leads to $t$ being a function of $\lambda$,
as we now proceed to show. The solution of Eq.\ (\ref{eq:3.7a}) that is
well-behaved for $\omega \rightarrow \infty$ can be written
\begin{eqnarray}
g_{\lambda}(\omega)&=&\left(\frac{\omega}{\lambda}\right)^{-\sqrt t}\,
   \frac{1}{(1 + \lambda/\omega)^{1+\sqrt{t}}}
\nonumber\\
&&\quad\ \times F\left(1+\sqrt{t},2+\sqrt{t};3;\frac{1}{1+\lambda/\omega}
                                     \right)\quad,
\label{eq:3.9}
\end{eqnarray}
with $F$ a hypergeometric function.
For later reference we also list the large-frequency behavior of the
original eigenfunction $f$, related to $g$ by Eq.\ (\ref{eq:3.3}).
By an analysis analogous to the one above for $g$ one finds
\begin{equation}
f_{\lambda}(\omega\rightarrow\infty) \propto (\omega/\lambda)^{-1-\sqrt{t}}
         \quad.
\label{eq:3.10}
\end{equation}

\subsection{The correlation length, and the exponents $\nu$, $\eta$, and
 $\gamma$}
\label{subsec:III.C}

Let us consider the eigenfunction, Eq.\ (\ref{eq:3.9}), in the limit
$\lambda\rightarrow 0$. Anticipating that $t\rightarrow 0$ as well in
that limit, we expand in $t$ and find asymptotically
\begin{equation}
g_{\lambda\rightarrow 0}(\omega) = -(\omega/\lambda)^{-\sqrt t}/\sqrt{t}\quad.
\label{eq:3.11}
\end{equation}
We know, however, that in this limit the eigenfunction must merge with the
solution $g_{\lambda=0}^{(1)}$, Eq.\ (\ref{eq:3.8}). This yields
$\lambda$ as a function of $t$ for asymptotically small $t$,
\begin{equation}
\lambda (t\rightarrow 0) \propto t^{1/2\sqrt{t}}\quad.
\label{eq:3.12}
\end{equation}
Furthermore, we know that at finite momentum ${\bf p}$, both the eigenvalue
$\lambda$ and ${\bf p}^2$ appear additively in the eigenvalue problem.
Scaling the momentum with a length scale $\xi$, we see that the instability
signaled by the appearance of a zero eigenvalue as $t\rightarrow 0$ is
accompanied by a diverging length scale
\begin{mathletters}
\label{eqs:3.13}
\begin{equation}
\xi(t\rightarrow 0) \propto ({\rm const.}\times t)^{-1/4\sqrt{t}}\quad.
\label{eq:3.13a}
\end{equation}
Clearly, $\xi$ is the correlation length for the phase transition that
is the result of the instability of the normal metal phase. According
to the usual definition of the correlation length exponent $\nu$,
$\xi \propto t^{-\nu}$, one has
\begin{equation}
\nu = \infty\quad.
\label{eq:3.13b}
\end{equation}
\end{mathletters}%

We now turn to the order parameter susceptibility and the critical
exponent $\gamma$. Since the diverging correlation length $\xi$ appears
in the particle-particle spin-triplet channel, we expect the order
parameter to be one of the corresponding components of the $Q$ matrix,
e.g. ${^1_1Q}$. The order parameter susceptibility will then be the
corresponding two-point correlation function. In the symbolic notation
of Sec.\ \ref{subsec:III.A}, the susceptibility is given by
\begin{eqnarray}
\chi(\Omega_n) &=& (\delta_n\vert M^{-1}\vert\delta_n)
\nonumber\\
    &\equiv& \sum_{1,2,3,4} \delta_{1+2,n}\,M^{-1}_{12,34}\,\delta_{3+4,n}
     \quad.
\label{eq:3.14}
\end{eqnarray}
Now consider the eigenvalue equation for the matrix $M$, Eq.\ (\ref{eq:3.2}),
and go into the basis of eigenfunctions $f^{(i)}$ to the eigenvalues
$\lambda_i$. Inserting two complete sets of eigenfunctions in 
Eq.\ (\ref{eq:3.14}) we obtain
\begin{equation}
\chi(\Omega_n) = \sum_i (\delta_n\vert f^{(i)})\,\frac{1}{\lambda_i\,
   (f^{(i)}\vert f^{(i)})}\,(f^{(i)}\vert \delta_n)\quad.
\label{eq:3.15}
\end{equation}
Let $\lambda$ be the smallest eigenvalue, and $f_{\lambda}$ the 
corresponding eigenfunction. Then the leading contribution to the
diverging susceptibility as $\lambda\rightarrow 0$ is obtained by keeping
only $f_{\lambda}$ in Eq.\ (\ref{eq:3.15}). At zero external frequency
we obtain
\begin{equation}
\chi(\Omega=0) = \left(\int_{-\infty}^{\infty}d\omega\,f_{\lambda}(\omega)
   \right)^2/\lambda\,\int_{-\infty}^{\infty} d\omega\,f^2_{\lambda}(\omega)
   \quad.
\label{eq:3.16}
\end{equation}
The divergence of $\chi$ as $\lambda\rightarrow 0$ is determined by the
ultraviolet behavior of the eigenfunction $f$. With Eq.\ (\ref{eq:3.10})
we obtain
\begin{mathletters}
\label{eqs:3.17}
\begin{equation}
\chi(\Omega=0) \propto t^{-1}\quad.
\label{eq:3.17a}
\end{equation}
This means that the critical exponent $\gamma$, defined by 
$\chi\propto t^{-\gamma}$, has its mean-field value,
\begin{equation}
\gamma = 1\quad.
\label{eq:3.17b}
\end{equation}
\end{mathletters}%

Finally, we determine the critical wavenumber dependence of the order
parameter susceptibility, i.e., the critical exponent $\eta$. Here the
disorder part of the action makes a difference, and so we consider it
explicitly. From Eqs.\ (\ref{eq:3.4b}) and (\ref{eq:2.15a}) we see
that, with $a={\cal D}^{-1}$, the eigenvalue $\lambda$ as a function of both
$t$ and the momentum ${\bf p}$ has the form 
$\lambda\propto t^{1/2\sqrt{t}} + {\rm const.}\times{\bf p}^2$. 
In a scaling sense,\cite{Notation_Footnote} we
therefore have $t^{1/2\sqrt{t}} \sim {\bf p}^2$, or
\begin{equation}
t \sim \left(\frac{\ln\ln 1/{\bf p}^2}{\ln 1/{\bf p}^2}\right)^2\quad,
\label{eq:3.18}
\end{equation}
plus terms that are less leading for ${\bf p}\rightarrow 0$. Together
with Eq.\ (\ref{eq:3.17a}) this implies for the critical susceptibility
\begin{mathletters}
\label{eqs:3.19}
\begin{equation}
\chi({\bf p},t=0) \propto \left(\frac{\ln 1/{\bf p}^2}{\ln\ln 1/{\bf p}^2}
                            \right)^2\quad.
\label{eq:3.19a}
\end{equation}
According to the definition of the critical exponent $\eta$,
$\chi({\bf p},t=0) \propto {\bf p}^{-(2-\eta)}$ this means
\begin{equation}
\eta = 2\quad,
\label{eq:3.19b}
\end{equation}
\end{mathletters}%
up to logarithmic corrections to power-law critical behavior.

\section{The ordered phase}
\label{sec:IV}

\subsection{The gap equation}
\label{subsec:IV.A}

We now turn to the ordered phase. Since the diverging susceptibility,
Sec.\ \ref{subsec:III.C}, is in the particle-particle spin triplet
channel, we know that the order parameter will be a $Q$ matrix in
that sector. The symmetry group in that sector is $U(1)\times SU(2)$
(the usual gauge symmetry plus rotations in spin space), which is
spontaneously broken to $Z_2\times U(1)$. We choose $r=s=1$ as the direction 
in which the symmetry is broken, and accordingly make an 
{\it ansatz}\cite{Frequency_Footnote}
\begin{mathletters}
\label{eqs:4.1}
\begin{eqnarray}
{_r^iQ}_{12}({\bf x})\Bigl\vert_{\rm sp}&=&\delta_{n_1,-n_2}
   \,\delta_{r1}\,\delta_{i1}\,Q_{n_1}\quad,
\label{eq:4.1a}\\
{_r^i\wt\Lambda}_{12}({\bf x})\Bigl\vert_{\rm sp}&=&\delta_{n_1,-n_2}\,
    \delta_{r1}\,\delta_{i1}\,\Lambda_{n_1}\quad,
\label{eq:4.1b}
\end{eqnarray}
\end{mathletters}%
for the saddle-point values of the fields in the broken-symmetry
phase. Using this {\it ansatz} in the saddle-point equations,
Eqs.\ (\ref{eq:2.8}), yields the following equations for $\Lambda_n$
and $Q_n$,
\begin{mathletters}
\label{eqs:4.2}
\begin{equation}
Q_n = -\frac{1}{2}\,\Lambda_{-n}\,\frac{1}{V}\sum_{\bf k}
      \frac{1}{\omega_n^2 + \xi_{\bf k}^2 - (\Lambda_n)^2}\quad,
\label{eq:4.2a}
\end{equation}
\begin{equation}
\Lambda_n = \frac{\pi T}{N_{\rm F}}\sum_m K_{nm}\,\frac{1}{V}\sum_{\bf k}
            \frac{\Lambda_m}{\omega_m^2 + \xi_{\bf k}^2 - (\Lambda_m)^2}
            \quad.
\label{eq:4.2b}
\end{equation}
Here
\begin{equation}
K_{nm} = y\,\ln\left\vert\frac{n-m}{n+m}\right\vert\quad,
\label{eq:4.2c}
\end{equation}
\end{mathletters}%
and we have used the fact that $\Lambda_n = -\Lambda_{-n}$ is an odd
function of the frequency.

It is clear from Eq.\ (\ref{eq:2.6a}) (see the third term on the r.h.s of
that equation) that $\Lambda_n$ plays the role of the self energy in the
particle-particle spin-triplet channel or of the gap function for the
triplet superconductor. A crucial question now arises about the
reality properties of $\Lambda_n$. It follows from Eq.\ (\ref{eq:4.2a})
that $\Lambda_n$ is real (imaginary) if and only if $Q_n$ is real (imaginary).
According to the formal symmetry properties, Eqs. (\ref{eq:2.5d}),
(\ref{eq:2.5e}), $Q_n$ should be real. However, as we have discussed at the
end of Sec.\ \ref{subsec:II.B}, this is misleading. In order for the field
theory to be stable and correctly reproduce perturbation theory, the
integration contour for the $Q$ must be deformed, and for our saddle point
to lie on the contour we must choose $Q_n$, and hence
$\Lambda_n$, to be imaginary. Another requirement is of course that this
choice minimizes the free energy. We will show in the next subsection that
this is indeed the case. We thus identify
$\Delta_n = -i\Lambda_n$, with $\Delta_n$ real,\cite{Gauge_Footnote}
as the gap function, and obtain the gap equation in the form
\begin{equation}
\Delta_n = \frac{\pi T}{N_{\rm F}}\sum_m K_{nm}\,\frac{1}{V}\sum_{\bf k}
            \frac{\Delta_m}{\omega_m^2 + \xi_{\bf k}^2 + (\Delta_m)^2}
            \quad.
\label{eq:4.3}
\end{equation}
Note that this gap equation has the same structure as the one in BCS theory
(except for the frequency dependences of the kernel and the gap function),
and that it is identical to the gap equation that was derived in
Ref.\ \onlinecite{us_tsc}.\cite{tsc_Footnote} It was shown in that
reference that there are nonzero solutions for $\Delta_n$ for $y>y_c=0.25$,
i.e. in the region in parameter space where we have found the normal
metal to be unstable. We also note that the gap equation is the same
irrespective of whether or not we explicitly keep the disorder term in
the action. This is the triplet analog of Anderson's theorem.

The gap equation, Eq.\ (\ref{eq:4.3}) was solved in Ref.\ \onlinecite{us_tsc},
both analytically in certain limits and numerically for all frequencies,
and there is no need to repeat this. Let us discuss, however, one important
point that was not mentioned in the earlier work. 
To this end, we first recall that by using the approximation, 
Eq.\ (\ref{eq:3.6}), for the kernel of the
integral equation, Eq.\ (\ref{eq:4.3}), one can transform the integral
equation into an ODE, as we did for the eigenfunction in 
Sec.\ \ref{subsec:III.B}. We briefly recall the most
important features of the solution. On the imaginary axis, 
$\Delta_n = \Delta (i\omega_n) \equiv \delta (\omega_n)$, and the limiting
behavior of $\delta (\omega)$ for asymptotically small $\omega$ is
\begin{mathletters}
\label{eqs:4.4}
\begin{equation}
\delta (\omega\rightarrow 0) = -2\,y\,\omega\ln (\omega/\omega_c)\quad,
\label{eq:4.4a}
\end{equation}
with $\omega_c$ a frequency scale that may depend on $y$, and for
asymptotically large frequencies
\begin{figure}[t]
\vskip 20mm
\centerline{\hskip 5mm\psfig{figure=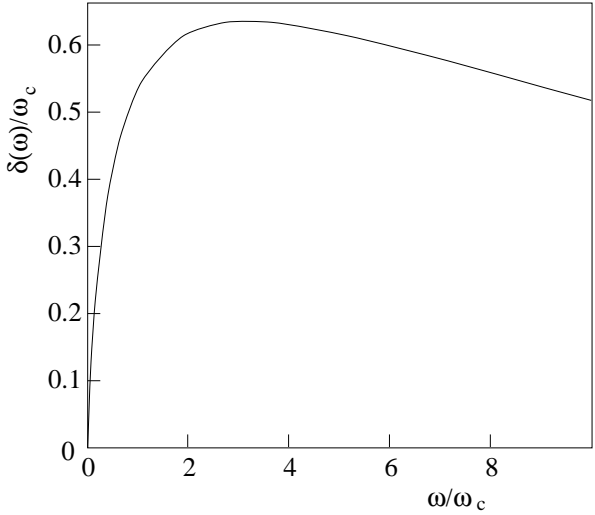,width=75mm}\vspace*{5mm}}
\vskip -15mm
\caption{Low-frequency behavior of the
 gap function for $y = 1.3\,y_c$. See the text for
 an explanation of the numerical procedure.}
\label{fig:1}
\end{figure}
\begin{figure}[t]
\vskip 17mm
\centerline {\hskip 13mm\psfig{figure=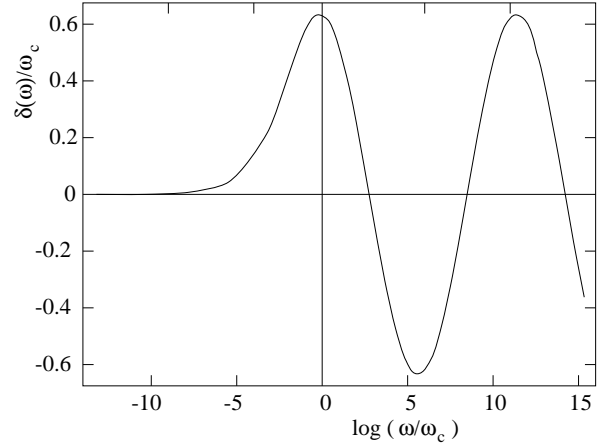,width=75mm}\vspace*{5mm}}
\vskip -12mm
\caption{Overall behavior of the
 gap function for $y = 1.3\,y_c$ on a logarithmic frequency
 scale. See the text for
 an explanation of the numerical procedure.}
\label{fig:2}
\end{figure}%
\begin{equation}
\delta (\omega\rightarrow\infty) = {\rm const.}\times
   \cos\left(\vert t\vert^{1/2}\ln\omega + \phi\right)\quad,
\label{eq:4.4b}
\end{equation}
\end{mathletters}%
with a constant prefactor and a phase $\phi$ that we could not determined
analytically. Figure\ \ref{fig:1} shows the function $\delta(\omega)$ for
small $\omega$, and Fig.\ \ref{fig:2} shows the oscillatory behavior 
at large $\omega$ on a
logarithmic scale. For this numerical solution at a fixed value of $y$,
Eq.\ (\ref{eq:4.4a}) was used to provide an initial condition for a very
small but non-zero frequency, and then the ODE was iterated towards larger
frequencies. For the initial condition, a value of $\omega_c$ was chosen
arbitrarily. It turns out that one finds a solution irrespective of the
value of $\omega_c$ chosen, and that all of these solutions are qualitatively
the same. This behavior of the numerics is readily understood by inspecting
the gap equation. It is easy to see that if $\delta (\omega)$ is a solution,
then so is $\gamma(\omega) = b^{-1}\,\delta(b\omega)$ with an arbitrary
positive number $b$. This scaling property holds for both the integral
equation, Eq.\ (\ref{eq:4.3}), and for the ODE that is derived from it
by approximating the kernel. This raises the question of what determines
the scale $\omega_c$ in Eq.\ (\ref{eq:4.4a}). In principle, the answer
must be inherent in the gap equation itself, but we can obtain $\omega_c$
from a much
simpler argument. As we have seen in Sec.\ \ref{sec:III}, the transition
is an ordinary continuous phase transition with a diverging length scale,
so scaling will work. We further know that frequencies scale like
wavenumbers squared, $\omega \sim {\bf p}^2$. From a scaling point of
view, $\omega_c$ is the critical frequency scale, and hence
\begin{equation}
\omega_c(t) \propto \xi^{-2}(t) \propto \vert t\vert^{1/2\sqrt{t}}\quad,
\label{eq:4.5}
\end{equation}
where we have used Eq.\ (\ref{eq:3.13a}). This means that 
$\delta_{t\rightarrow 0}(\omega) \rightarrow 0$ for all $\omega$, as one
would expect, and the overall amplitude of the order parameter vanishes
like $\omega_c$. For the critical exponent $\beta$, defined by
$\Delta \propto t^{\beta}$, this implies
\begin{equation}
\beta = 2\nu = \infty\quad.
\label{eq:4.6}
\end{equation}

\subsection{The free energy}
\label{subsec:IV.B}

We now check whether the saddle-point solution that we found to exist
is stable, i.e. whether it has a free energy that is lower than that
of the normal metal state. In mean-field approximation, the
free energy is given by the action, with the fields
replaced by their mean-field values. By substituting Eqs.\ (\ref{eqs:4.1})
in Eq.\ (\ref{eq:2.6a}) and expanding and resumming the $\Tr\ln$ term,
we find for the free energy density in mean-field approximation
\begin{equation}
f = -2T\sum_n \Lambda_n\,Q_n - T\sum_n\frac{1}{V}\sum_{\bf k}
    \ln\left[1\! -\! \Lambda_n^2/(\omega_n^2 + \xi_{\bf k}^2)\right],
\label{eq:4.7}
\end{equation}
where we have used Eqs.\ (\ref{eqs:4.2}) to rewrite the term that was
quadratic in $Q_n$ as a term that is bilinear in $Q_n$ and $\Lambda_n$.
We next use Eqs.\ (\ref{eqs:4.2}) again to make this term quadratic in
$\Lambda_n$,
\begin{mathletters}
\label{eqs:4.8}
\begin{eqnarray}
f &=& \frac{-N_{\rm F}}{\pi}\,T\sum_{n,m}\Lambda_n\,K^{-1}_{nm}\,\Lambda_m
\nonumber\\
    &&-T\sum_n\frac{1}{V}\sum_{\bf k}
      \ln\left[1 - \Lambda_n^2/(\omega_n^2 + \xi_{\bf k}^2)\right]\quad,
\label{eq:4.8a}
\end{eqnarray}
where $K^{-1}$ is the inverse of the operator $K$, Eq.\ (\ref{eq:4.2c}),
in the sense that
\begin{equation}
T\sum_{n'} K^{-1}_{nn'}\,K_{n'm} = \delta_{nm}\quad.
\label{eq:4.8b}
\end{equation}
\end{mathletters}%
Now we use $\Lambda_n = -i\Delta_n$, and expand in powers of the order
parameter. This yields the Landau expansion
\begin{mathletters}
\label{eqs:4.9}
\begin{equation}
f = T^2\sum_{n,m} \Delta_n\,t_{nm}\,\Delta_m + T\sum_n u_n\,\Delta^4_n
     + O(\Delta^6)\quad,
\label{eq:4.9a}
\end{equation}
where
\begin{equation}
t_{nm} = \frac{N_{\rm F}}{\pi T}\,K^{-1}_{nm} - \frac{1}{T}\,\delta_{nm}\,
         \frac{1}{V}\sum_{\bf k}\frac{1}{\omega_n^2 + \xi_{\bf k}^2}\quad,
\label{eq:4.9b}
\end{equation}
and
\begin{equation}
u_n = \frac{1}{2V}\sum_{\bf k}\frac{1}{(\omega_n^2 + \xi_{\bf k}^2)^2}>0\quad.
\label{eq:4.9c}
\end{equation}
\end{mathletters}%
$u_n$ is positive definite, and hence the theory is stable. Furthermore,
$t_{nm}$ is negative for small $K^{-1}$ or large $K$, i.e. for strong
coupling. Hence the free energy in a state with a nonzero order parameter
is indeed lower than in a state with zero order parameter, provided that
the coupling is sufficiently strong, i.e. $y>y_c$. All of these considerations
are in exact analogy to the case of BCS theory. We note that if $\Lambda_n$
were real, then the sign of the quadratic term in Eq.\ (\ref{eq:4.9a})
would be reversed, and consequently the free energy would favor order
at {\it small} coupling, which is somewhat paradoxical. This was actually
mentioned already in Berezinskii's paper,\cite{Berezinskii} and later
investigated in more detail by Heid.\cite{Heid}

\subsection{Linear response in the ordered phase}
\label{subsec:IV.C}

We now calculate the linear response in the ordered phase, specifically
the magnetic susceptibility and the electrical conductivity, in Gaussian
approximation. This will establish that the ordered phase is really a
superconductor.

To start with, we need the saddle-point Green function. It's inverse
is (see Eq.\ (\ref{eq:2.6a}))
\begin{equation}
G^{-1} = G_0^{-1} - i{\wt \Lambda}\quad.
\label{eq:4.10}
\end{equation}
Using Eq.\ (\ref{eq:4.1b}), and inverting the $2\times 2$ matrix, we
obtain
\begin{mathletters}
\label{eqs:4.11}
\begin{equation}
G_{nm}({\bf k}) = \delta_{nm}\,{\cal G}_n({\bf k})\,
                         (\tau_0\otimes s_0) 
   + \delta_{n,-m}\,{\cal F}_n ({\bf k})\,(\tau_1\otimes s_1)\ ,
\label{eq:4.11a}
\end{equation}
where
\begin{equation}
{\cal G}_n({\bf k}) = \frac{-(i\omega_n + \xi_{\bf k})}{\omega_n^2 
                             + \xi_{\bf k}^2 + \Delta_n^2}\quad,
\label{eq:4.11b}
\end{equation}
\begin{equation}
{\cal F}_n({\bf k}) = \frac{-\Delta_n}{\omega_n^2 + \xi_{\bf k}^2 + \Delta_n^2}
                                                            \quad.
\label{eq:4.11c}
\end{equation}
\end{mathletters}%
Notice again the structural analogy to BCS theory.

\subsubsection{The anomalous susceptibility}
\label{subsubsec:IV.B.1}

Before we calculate observable correlation functions, let us focus on
the Goldstone modes. Although not directly observable, they are of crucial
importance for the former. As we already noted in Sec.\ \ref{subsec:IV.A},
the relevant symmetry group is $U(1)\times SU(2)$, which is spontaneously
broken to the little group $Z_2\times U(1)$. The dimension of the relevant
quotient space is ${\rm dim}\,\left(U(1)\times SU(2)/U(1)\right) = 5$,
so there are five Goldstone modes. With our choice of the order parameter
in the sector $r=s=1$, they are represented by the correlation functions
\begin{equation}
T\sum_{1,2} \left\langle{^i_rQ}_{1+n,-1}({\bf k})\,{^i_rQ}_{2+n,-2}(-{\bf k})
                  \right\rangle\quad,
\label{eq:4.12}
\end{equation}
with $r=1; i=1,2$, or $r=2; i=1,2,3$. It will turn out that the $r=2,i=1$
Goldstone mode couples to the density response, so let us focus on the
anomalous susceptibility
\begin{equation}
\chi^{(a)}_{12,34}({\bf k}) = \left\langle{^1_2Q}_{1,2}({\bf k})\,{^1_2Q}_{3,4}
                      (-{\bf k})\right\rangle\quad.
\label{eq:4.13}
\end{equation}
Now consider the eigenvalue problem
\begin{equation}
\sum_{3,4}\chi^{(a)}_{12,34}\,f_{34} = \lambda\,f_{12}\quad.
\label{eq:4.14}
\end{equation}
This is the same eigenvalue problem as the one we considered in 
Sec.\ \ref{sec:III} in the symmetric phase. We therefore know that at
zero momentum there is an eigenvalue $\lambda_0$ that is infinite everywhere 
in the broken symmetry phase. 
We further know (see Sec.\ \ref{subsec:III.C}) that the wavenumber dependence
of $\lambda^{-1}$ quadratic. Perturbation theory can be used to convince oneself
that the frequency dependence is also quadratic, although determining the
prefactors would be hard. Dropping all prefactors, we thus have the
structure
\begin{equation}
\lambda({\bf p},\Omega) = \frac{1}{{\bf p}^2 + \Omega^2}\quad,
\label{eq:4.15}
\end{equation}
everywhere in the ordered or broken symmetry phase.

\subsubsection{The transverse current susceptibility}
\label{subsubsec:IV.B.2}

Let us now calculate the transverse current susceptibility, which determines
the magnetic susceptibility and hence the Meissner effect. By adding
an appropriate source term to our action, we obtain for the transverse
susceptibility in saddle-point approximation
\begin{eqnarray}
\chi_{\rm T}({\bf k}\rightarrow 0) &=& \frac{1}{m^2}\sum_{\bf p} {\bf p}^2\,T
   \sum_n \left[-{\cal G}_n({\bf p})\,{\cal G}_n({\bf p})\right.
\nonumber\\
 &&\qquad\qquad\qquad\left. + {\cal F}_n({\bf p})\,{\cal F}_{-n}({\bf p})
                                                               \right]\quad,
\label{eq:4.16}
\end{eqnarray}
in $D=2$ dimensions. Notice that in a diagrammatic language, 
Eq.\ (\ref{eq:4.16}) is a simple bubble. In the transverse channel, this is
sufficient since all vertex corrections vanish. In the language of the
present field theory, saddle-point and Gaussian approximations coincide.
Now we do the integrals
in Eq.\ (\ref{eq:4.16}). A convenient way to do this is to put an upper
cutoff on the integral over $\xi_{\bf p}$, do that integral first, and let
the cutoff go to infinity in the end. One can readily check that this
procedure yields the correct answer in the BSC case (where convergence
problems can be avoided by doing the frequency integration first).
We then find\cite{us_mosfets}
\begin{equation}
\chi_{\rm T}({\bf k}\rightarrow 0) = \frac{n}{m}\,\left[1 - 
   \int_{0}^{\infty} d\omega\,\frac{\delta^2(\omega)}{\left(\omega^2 + \delta^2
     (\omega)\right)^{3/2}}\right]\quad.
\label{eq:4.17}
\end{equation}
The integral is positive definite, therefore $\chi_{\rm T} < n/m$ and we
have an ordinary Meissner effect. Notice that the reality properties of
$\Lambda$ and $\Delta$ are once again crucial for this conclusion: If
$\Lambda$ in the particle-particle spin-triplet channel were real, then
the correction to the free fermions result in Eq.\ (\ref{eq:4.16}) would
be positive! We also note that $\delta(\omega)$ as given by
Eq.\ (\ref{eq:4.4a}) renders the frequency integral in Eq.\ (\ref{eq:4.16})
infinite. This is an artifact of the theory that is due to a missing
self-consistency. As was discussed in Ref.\ \onlinecite{us_mosfets},
taking into account the feedback of the nonzero order parameter on
the electron-electron interaction would cure this problem.

\subsubsection{The longitudinal response}
\label{subsubsec:IV.B.3}

The calculation of the longitudinal current, or density, response, and
hence the conductivity, is much more involved. The reason is that in order
to obtain a conserving approximation, one needs to solve the full Gaussian
theory (in diagrammatic language, one needs to take the vertex corrections
into account). This is well known in BCS theory,\cite{AndersonNambu} and
the calculation for the present case proceeds analogously. As a check of
our procedure, we have also performed the calculation for the BCS case
and have reproduced the standard results. 

We start out by expanding the action, Eq.\ (\ref{eq:2.6a}), to Gaussian
order about the saddle-point solution, Eqs.\ (\ref{eqs:4.2}). For the
part that is quadratic in 
$\delta{\wt\Lambda} = {\wt\Lambda} - {\wt\Lambda}_{\rm sp}$
we find
\begin{mathletters}
\label{eqs:4.18}
\begin{eqnarray}
\Tr\left(G_{\rm sp}\,\delta{\wt\Lambda}\,G_{\rm sp}\,\delta{\wt\Lambda}\right)
  &=& 4\sum_{1,2,3,4}\frac{1}{V}\sum_{\bf k}\sum_{r,s}\sum_{i,j}
      {^i_r(}\delta{\wt\Lambda})_{12}({\bf k})
\nonumber\\
&&\times\ {^{ij}_{rs}{\wt A}}_{12,34}({\bf k})\,
          {^j_s(}\delta{\wt\Lambda})_{34}(-{\bf k})\quad.
\nonumber\\
\label{eq:4.18a}
\end{eqnarray}
Here
\begin{eqnarray}
{^{ij}_{rs}{\wt A}}_{12,34}({\bf k}) &=& \delta_{13}\,\delta_{24}\,
                       \varphi^{00}_{12}({\bf k})\,M^{00}_{rs}\,\delta_{ij}
\nonumber\\
 && + \delta_{13}\,\delta_{2,-4}\,\varphi^{01}_{12}({\bf k})\,M^{01}_{rs}\,
                                                              m^{01}_{ij}
\nonumber\\
 && + \delta_{1,-3}\,\delta_{24}\,\varphi^{10}_{12}({\bf k})\,M^{10}_{rs}\,
                                                              m^{10}_{ij}
\nonumber\\
 && + \delta_{1,-4}\,\delta_{2,-3}\,\varphi^{11}_{12}({\bf k})\,M^{11}_{rs}\,
                                                          M^{11}_{ij}\quad,
\label{eq:4.18b}
\end{eqnarray}
with $4\times 4$ matrices
\begin{eqnarray}
M^{00}&=&\left(\begin{array}{cc} i\tau_3 & 0 \\
                                    0    & -i\tau_3\\ \end{array}\right)\quad,
\quad
M^{01} = \left(\begin{array}{cc} -i\tau_1 & 0 \\
                                    0     & -i\tau_1\\ \end{array}\right)\quad,
\nonumber\\
M^{10}&=&\left(\begin{array}{cc} -i\tau_1 & 0 \\
                                    0     & i\tau_1\\ \end{array}\right)\quad,
\quad
M^{11} = \left(\begin{array}{cc} -i\tau_3 & 0 \\
                                    0     & -\tau_0\\ \end{array}\right)\quad,
\nonumber\\
m^{01}&=&\left(\begin{array}{cc} \tau_2 & 0 \\
                                    0     & -\tau_2\\ \end{array}\right)
\quad\ \ ,
\quad\ 
m^{10} = \left(\begin{array}{cc} \tau_2 & 0 \\
                                    0     & \tau_2\\ \end{array}\right)\quad.
\nonumber\\
\label{eq:4.18c}
\end{eqnarray}
and
\begin{equation}
\varphi^{00}_{nm}({\bf k}) = \frac{1}{V}\sum_{\bf p} {\cal G}_n({\bf p})\,
        {\cal G}_m({\bf p} + {\bf k})\quad,
\label{eq:4.18i}
\end{equation}
\end{mathletters}%
and $\varphi^{01}$, $\varphi^{10}$, and $\varphi^{11}$ defined analogously
with ${\cal G}{\cal G}$ in Eq.\ (\ref{eq:4.18i}) replaced by 
${\cal G}{\cal F}$, ${\cal F}{\cal G}$, and ${\cal F}{\cal F}$, respectively.

Similarly, the term that couples $\delta{\wt\Lambda}$ and $\delta Q$ can
be written
\begin{mathletters}
\label{eqs:4.19}
\begin{eqnarray}
\Tr\left(\delta{\wt\Lambda}\,\delta Q\right)
  &=& 4\sum_{1,2,3,4}\frac{1}{V}\sum_{\bf k}\sum_{r,s}\sum_{i,j}
      {^i_r(}\delta{\wt\Lambda})_{12}({\bf k})
\nonumber\\
&\times&{^{ij}_{rs}{\wt B}}_{12,34}({\bf k})\,
          {^j_s(}\delta Q)_{34}(-{\bf k})\ ,
\label{eq:4.19a}
\end{eqnarray}
where
\begin{equation}
{^{ij}_{rs}{\wt B}}_{12,34}({\bf k}) = \delta_{13}\,\delta_{24}\,\delta_{rs}\,
  \left({{+\atop -}\atop{-\atop +}}\right)_r\,\delta_{ij}\,
  \left({{+\atop -}\atop{-\atop -}}\right)_i\quad.
\label{eq:4.19b}
\end{equation}
\end{mathletters}%
The full Gaussian action is
\begin{equation}
{\cal A}_{\rm G} = \Tr\left(G_{\rm sp}\,\delta{\wt\Lambda}\,G_{\rm sp}\,
   \delta{\wt\Lambda}\right)
+ \Tr\left(\delta{\wt\Lambda}\,\delta Q\right)
+ {\cal A}_{\rm int}^{\rm p-p,t}[\delta Q]\ .
\label{eq:4.20}
\end{equation}
 
This Gaussian action is too complicated to be handled conveniently, and
approximations are necessary. In order to make sensible approximations, we
recall that our purpose is to calculate the density response in a conserving 
approximation. In order to do so, it is crucial to preserve the structure
of the soft modes in the theory, while massive modes can be dealt with in
very crude approximations. It turns out that the soft mode structure is
preserved by keeping only the imaginary parts of the fields in the
$r=1,2,3$ channels, and only the real parts in the $r=0$ channel, and
integrating out the remaining parts in saddle-point approximation.
This amounts to making the approximations
\begin{mathletters}
\label{eqs:4.21}
\begin{eqnarray}
{^i_r(}\delta{\wt\Lambda})_{12} &\approx& \frac{1}{2}\,\left[
 {^i_r(}\delta{\wt\Lambda})_{12} + \left({{+\atop +}\atop{+\atop -}}\right)_r\,
     {^i_r(}\delta{\wt\Lambda})_{-1,-2}\right]
\nonumber\\
&=& \left({{i{\rm Im}\atop i{\rm Im}}\atop{i{\rm Im}\atop{\rm Re}}}\right)_r\,
     {^i_r(}\delta{\wt\Lambda})_{12} \equiv {^i_r\lambda}_{12}\quad,
\label{eq:4.21a}
\end{eqnarray}
and approximating $\delta Q$ by an analogously defined object $q$.
$\lambda$ and $q$ have the symmetry properties
\begin{equation}
{^i_r\lambda}_{12} = \left({{+\atop +}\atop{+\atop -}}\right)_r\,
                                    {^i_r\lambda}_{-1,-2}\quad,
\label{eq:4.21b}
\end{equation}
\begin{equation}
{^i_r q}_{12} = \left({{+\atop +}\atop{+\atop -}}\right)_r\, 
                                    {^i_r q}_{-1,-2}\quad,
\label{eq:4.21c}
\end{equation}
\end{mathletters}%
The net effect of these approximations is a symmetrization of the action
with respect to positive and negative frequency values. 

We now formally integrate out $\delta{\wt\Lambda}$. This yields the
action entirely in terms of $q$,
\begin{mathletters}
\label{eqs:4.22}
\begin{eqnarray}
{\cal A}_{\rm G}[q] &=& \frac{-4}{V}\sum_{\bf p}\sum_{1,2,3,4}
   \sum_{r,s}\sum_{i.j}
   {^i_r q}_{12}({\bf p})\,
\nonumber\\
&&\times\left[{^{ij}_{rs}A}^{-1}_{12,34}({\bf p}) - {^{ij}_{rs}B}_{12,34}
        \right]\,{^j_s q}_{34}(-{\bf p})\quad.
\label{eq:4.22a}
\end{eqnarray}
Here ${^{ij}_{rs}A}^{-1}_{12,34}$ is the inverse of the matrix
\begin{eqnarray}
{^{ij}_{rs}A}_{12,34} &=& \frac{1}{4}\,
\left({{+\atop -}\atop{-\atop -}}\right)_i\,
        \left({{+\atop -}\atop{-\atop -}}\right)_j
\left[
   \left({{+\atop -}\atop{-\atop +}}\right)_r\,
      \left({{+\atop -}\atop{-\atop +}}\right)_s\,{^{ij}_{rs}{\wt A}}_{12,34}
               \right.
\nonumber\\
&& + \left.\left({{+\atop -}\atop{-\atop -}}\right)_r\,
    \left({{+\atop -}\atop{-\atop +}}\right)_s\,{^{ij}_{rs}{\wt A}}_{-1,-2;3,4}
             \right.
\nonumber\\
&& + \left({{+\atop -}\atop{-\atop +}}\right)_r\,
    \left({{+\atop -}\atop{-\atop -}}\right)_s\,
       {^{ij}_{rs}{\wt A}}_{1,2;-3,-4}
\nonumber\\
&&  + \left.\left({{+\atop -}\atop{-\atop -}}\right)_r\,
       \left({{+\atop -}\atop{-\atop -}}\right)_s\,
            {^{ij}_{rs}{\wt A}}_{-1,-2;-3,-4}\right]\ ,
\label{eq:4.22b}
\end{eqnarray}
and
\begin{equation}
{^{ij}_{rs}B}_{12,34} = \delta_{rs}\,\delta_{ij}\,
   \left({{0\atop +}\atop{+\atop 0}}\right)_r\,
      \left({{0\atop +}\atop{+\atop +}}\right)_i\,B_{12,34}\quad,
\label{eq:4.22c}
\end{equation}
with
\begin{equation}
B_{12,34} = -\frac{1}{4}\,\pi N_{\rm F}\,T\,\delta_{1+2,3+4}\,
            {\wt K}_{12,34}\quad.
\label{eq:4.22d}
\end{equation}
\end{mathletters}%

By adding an
appropriate source term to the action it is easily checked that the
expression for the density susceptibility $\chi$ in terms of the $q$
is the same as in terms of the $Q$, viz.
\begin{mathletters}
\label{eqs:4.23}
\begin{equation}
\chi({\bf k},\Omega_n) = 16T\sum_{1,2} \sum_{r=0,3}
   \left\langle{^0_r q}_{1+n,1}({\bf k})\,{^0_r q}_{2+n,2}(-{\bf k})
            \right\rangle\quad.
\label{eq:4.23a}
\end{equation}
From the expression of $Q$ in terms of the fermion fields, Eq.\ (\ref{eq:2.2}),
it is easy to see that the contributions to Eq.\ (\ref{eq:4.23a}) from
$r=0$ and $r=3$ are identical, except at zero external frequency.
We can therefore write
\begin{equation}
\chi({\bf k},\Omega_n\neq 0) = 32T\sum_{1,2}
       \left\langle{^0_3 q}_{1+n,1}({\bf k})\,{^0_3 q}_{2+n,2}(-{\bf k})
            \right\rangle\quad,
\label{eq:4.23b}
\end{equation}
\end{mathletters}%
and obtain the zero frequency susceptibility from Eq.\ (\ref{eq:4.23b}) in
the limit $\Omega_n\rightarrow 0$.\cite{ConservingApproximation_Footnote}
Combining Eqs.\ (\ref{eq:4.23b}) and (\ref{eq:4.22a}) we have
\begin{mathletters}
\label{eqs:4.24}
\begin{equation}
\chi({\bf k},\Omega_n\neq 0) = 4T\sum_{1,2}
   {^{00}_{33}M}^{-1}_{1+n,1;2+n,2}({\bf k})\quad,
\label{eq:4.24a}
\end{equation}
where
\begin{equation}
M^{-1} = \left(A^{-1} - B\right)^{-1} \quad.
\label{eq:4.24b}
\end{equation}
\end{mathletters}%

Instead of inverting the matrix $M$ directly, it is convenient to rewrite
Eq.\ (\ref{eq:4.24b}) as an integral equation,
\begin{equation}
M^{-1} = A + A\,B\,M^{-1}\quad.
\label{eq:4.25}
\end{equation}
Ignoring the frequency indices for the time being, we need the matrix
element 
\begin{equation}
{^{00}_{33}M}^{-1} = {^{00}_{33}A} + {^{01}_{32}A}\,B\,{^{10}_{23}M}^{-1}
                          \quad,
\label{eq:4.26}
\end{equation}
where we have used the structure of $B$, Eq.\ (\ref{eq:4.22c}).
${^{10}_{23}M}^{-1}$ in turn obeys the integral equation
\begin{mathletters}
\label{eqs:4.27}
\begin{eqnarray}
{^{10}_{23}M}^{-1} &=& {^{10}_{23}A} + {^{11}_{22}A}\,B\,{^{10}_{23}M}^{-1}
\nonumber\\
                   &=& {^{10}_{23}A} + {^{11}_{22}A}\,\beta\,{^{10}_{23}A}
                           \quad,
\label{eq:4.27a}
\end{eqnarray}
where $\beta$ is the solution of
\begin{equation}
\beta = B + B\,{^{11}_{22}A}\,\beta\quad.
\label{eq:4.27b}
\end{equation}
$\beta$ is related to the anomalous susceptibility $\chi^{(a)}$,
Eq.\ (\ref{eq:4.13}), by
\begin{equation}
\beta = {^{11}_{22}A}^{-1}\,\chi^{(a)}\,B\quad,
\label{eq:4.27c}
\end{equation}
\end{mathletters}%
with our approximation for $\chi^{(a)}$ that replaces $Q$ in 
Eq.\ (\ref{eq:4.13}) by $q$. 

We are now in a position to calculate the density susceptibility.
Defining a vector $\vert\delta_n) \equiv \delta_{1,2-n}$ we can
write $\chi$, Eq.\ (\ref{eq:4.24a}), as a matrix element
\begin{equation}
\chi({\bf k},\Omega_n\neq 0) = 4T\,(\delta_n\vert\,{^{00}_{33}M}^{-1}({\bf k})
   \,\vert\delta_n)\quad.
\label{eq:4.28}
\end{equation}
Let us first consider the `bubble contribution',
\begin{mathletters}
\label{eqs:4.29}
\begin{equation}
\chi_0 ({\bf k},\Omega_n) = 4T\,(\delta_n\vert\,{^{00}_{33}A}({\bf k})\,
     \vert\delta_n)\quad.
\label{eq:4.29a}
\end{equation}
Equations (\ref{eq:4.22b}) and (\ref{eqs:4.18}) yield
\begin{eqnarray}
\chi_0 ({\bf k},\Omega_n) &=& T\sum_1 \left[\left(\varphi^{00}_{1+n,1}({\bf k})
   - \varphi^{11}_{1+n,1}({\bf k})\right)\right.
\nonumber\\
&&\qquad\qquad + \left.\left(\varphi^{00}_{1-n,1}({\bf k})
   - \varphi^{11}_{1-n,1}({\bf k})\right)\right]
\nonumber\\
&\equiv& I_1({\bf k},\Omega_n)\quad.
\label{eq:4.29b}
\end{eqnarray}
Doing the integrals in the limit of small wavenumber and frequency yields
\begin{equation}
I_1({\bf k},\Omega_n) = N_{\rm F} + O({\bf k}^2,\Omega)\quad.
\label{eq:4.29c}
\end{equation}
\end{mathletters}%
Notice that the bubble contribution is not a conserving approximation for
$\chi$, as the susceptibility does not vanish at zero wavenumber. As in
the BCS case, particle number conservation or gauge invariance will be
restored by the coupling of the density fluctuations to a soft or massless
Goldstone mode, as we now proceed to show.

We now consider the `soft' part of $\chi$, i.e. the one that couples to
the Goldstone mode. From Eqs.\ (\ref{eq:4.28}), (\ref{eqs:4.27}) we
find
\begin{mathletters}
\label{eqs:4.30}
\begin{eqnarray}
\chi_{\rm soft}({\bf k},\Omega_n)&=&(\delta_n\vert\,
  {^{01}_{32}A}({\bf k})\,B\,\chi^{(a)}({\bf k})\,B\,
     {^{10}_{23}A}({\bf k})\vert\delta_n)
\nonumber\\
&=&\sum_{f,f'} (\delta_n\vert\,{^{01}_{32}A}({\bf k})\,B\vert f)\,
     (f\vert\chi^{(a)}({\bf k})\vert f')\,
\nonumber\\
&&\qquad\qquad\quad\times (f'\vert B\,{^{10}_{23}A}({\bf k})\vert\delta_n)
                          \quad,
\label{eq:4.30a}
\end{eqnarray}
where we have inserted two complete sets $\{\vert f)\}$ and $\{\vert f')\}$
of eigenfunctions of $\chi^{(a)}$. Since we are interested in the coupling
to the soft mode, we need to keep only the eigenfunction $f$ whose eigenvalue
$\lambda$ is infinite at zero momentum and frequency, see Eq.\ (\ref{eq:4.15}).
We thus have
\begin{equation}
\chi_{\rm soft}({\bf k},\Omega_n) = (\delta_n\vert\,{^{01}_{32}A}({\bf k})
  \vert g)\,\lambda({\bf k},\Omega_n)\,(g\vert{^{10}_{23}A}({\bf k})\vert
   \delta_n)\quad,
\label{eq:4.30b}
\end{equation}
\end{mathletters}%
with $\vert g) = B\vert f)$ the transformed eigenfunction from 
Eq.\ (\ref{eq:3.3}). The structure of the eigenvalue we know, 
Eq.\ (\ref{eq:4.15}), and what remains to be done is to calculate the coupling
integral. Using the symmetry property $g_{12} = -g_{21}$ of the eigenfunction
that follows from Eq.\ (\ref{eq:3.4a}), we find
\begin{mathletters}
\label{eqs:4.31}
\begin{eqnarray}
I_2({\bf k},\Omega_n)&\equiv&(\delta_n\vert\,{^{01}_{32}A}({\bf k})\vert g)
   = (g\vert{^{10}_{23}A}({\bf k})\vert\delta_n)
\nonumber\\
&=&\frac{1}{2}\sum_1\left[\left(\varphi^{01}_{1+n,1}\,g_{1+n,-1} 
                             - \varphi^{01}_{1-n,1}\,g_{1-n,-1}\right)\right.
\nonumber\\
  &&\qquad + \left. \left(\varphi^{01}_{1+n,1}\,g_{1+n,-1}
        - \varphi^{01}_{1-n,1}\,g_{1-n,-1}\right)\right] \ .
\nonumber\\
\label{eq:4.31a}
\end{eqnarray}
Using the above symmetry property of $g$, and the fact that the anomalous
Green function ${\cal F}$, Eq.\ (\ref{eq:4.11c}), is an antisymmetric
function of the frequency, we see that $I_2$ is also an odd function of
its frequency argument. A Taylor expansion yields
\begin{equation}
I_2({\bf k},\Omega_n) = -I_2({\bf k},-\Omega_n) \propto \Omega_n\quad.
\label{eq:4.31b}
\end{equation}
\end{mathletters}%
We conclude that the density susceptibility at small frequencies and
wavenumbers has the structure
\begin{mathletters}
\label{eqs:4.32}
\begin{equation}
\chi({\bf k},\Omega_n) = N_{\rm F}\,\left(1 - \frac{a\,\Omega_n^2}
    {b\,{\bf k}^2 + c\,\Omega_n^2}\right)\quad,
\label{eq:4.32a}
\end{equation}
where $a,b,c$ are constants that we have not determined explicitly.
Since only the coupling to the Goldstone mode can restore the
property $\chi({\bf k}\rightarrow 0,\Omega_n) = 0$, we must have
$a=c$. We have performed the analogous procedure for the
BCS case, where $a=c$ is easily confirmed 
explicitly. We conclude that the structure of the density susceptibility
is the same as in the BCS case, viz.
\begin{equation}
\chi({\bf k},\Omega_n) = N_{\rm F}\,\frac{(v{\bf k})^2}{\Omega_n^2
   + (v{\bf k})^2}\quad,
\label{eq:4.32b}
\end{equation}
\end{mathletters}%
with $v$ the velocity of the sound-like mode that is the analog of the
Anderson-Bogolubov mode in BCS superconductors.\cite{ABModeFootnote}
Here we have not determined $v$ explicitly.

The electrical conductivity $\sigma$ is determined by $\chi$ via
\begin{equation}
\sigma(\Omega_n) = e^2\lim_{{\bf k}\rightarrow 0}\,\frac{\Omega_n}{{\bf k}^2}\,
   \chi({\bf k},\Omega_n)\quad.
\label{eq:4.33}
\end{equation}
Combining Eqs.\ (\ref{eq:4.32b}) and (\ref{eq:4.33}), we see that
$\sigma(\Omega_n) \propto 1/\Omega_n$, and the real part of the
conductivity as a function of real frequencies has a delta-function
contribution
\begin{equation}
{\rm Re}\ \sigma(\Omega) = \frac{e^2\,N_{\rm F}\,v^2}{\pi}\,
                              \delta(\Omega)\quad.
\label{eq:4.34}
\end{equation}
The ordered phase is thus a true superconductor. If we neglect the explicit
disorder term in the action, then the delta function in Eq.\ (\ref{eq:4.34})
will exhaust the f-sum rule, which in turn determines the velocity $v$.
However, with the disorder term taken into account, there is a continuous
spectrum in addition to the delta function, and the determination of $v$
would require an explicit calculation of the prefactors in
Eq.\ (\ref{eq:4.32a}).

\section{Discussion and Conclusion}
\label{sec:V}

\subsection{Summary of results}
\label{subsec:V.A}

Let us summarize. We have considered in detail the consequences of the
attractive interaction in the particle-particle spin-triplet channel
that has been predicted to exist in $2$-$D$ disordered interacting electron
systems.\cite{us_tsc} We have found that this interaction leads to
an instability of any normal metal phase in the system. The
instability is characterized by a diverging length scale $\xi$, and
has all the characteristics of a continuous quantum phase transition.
The critical exponents $\nu$, $\gamma$, and $\eta$ are
\begin{mathletters}
\label{eqs:5.1}
\begin{equation}
\nu=\infty\quad,\quad\gamma=1\quad,\quad\eta=2\quad,
\label{eq:5.1a}
\end{equation}
where the infinite value of $\nu$ indicates an essential singularity in
the dependence of $\xi$ on the distance from criticality $t$,
\begin{equation}
\xi \propto (c\,t)^{-1/4\sqrt{t}}\quad,
\label{eq:5.1b}
\end{equation}
\end{mathletters}%
with $c$ a constant, and the value of $\eta$ is to be understood
as a complicated logarithmic dependence of the critical order parameter
susceptibility on the wavenumber, Eq.\ (\ref{eq:3.19a}).

This instability identifies the order parameter of the phase transition,
viz. an anomalous density in the particle-particle spin-triplet channel.
Approaching the transition from the ordered phase, the order parameter
critical exponent is
\begin{mathletters}
\label{eqs:5.2}
\begin{equation}
\beta = 2\nu\quad,
\label{eq:5.2a}
\end{equation}
in the sense that the amplitude of the frequency dependent order parameter 
vanishes like
\begin{equation}
\Delta \propto \xi^{-2}\quad.
\label{eq:5.2b}
\end{equation}
\end{mathletters}%
The phase transition thus 
has all the characteristics of an ordinary quantum critical point.
The ordered phase has all the characteristics of a superconductor,
including a Meissner effect and an infinite static electrical conductivity.
While these conclusions are general, the detailed critical behavior, and
in particular the values of the exponents, depend on our particular
choice of the interaction kernel, Eq.\ (\ref{eq:2.7c}).

\subsection{Discussion}
\label{subsec:V.B}

As we have mentioned before, the reality properties of the gap function
are crucial for the stability of our superconducting state. If the
expectation value of ${^1_1Q}$ were real, then the sign of the
$(\Delta_m)^2$ term in Eq.\ (\ref{eq:4.3}) would be negative.
Irrespective of whether or not this modified gap equation has
physical solutions, this would lead to a free energy that is minimized
by a non-zero gap function at {\em weak} coupling. This was noticed
already by Berezinskii,\cite{Berezinskii} who proposed that odd-gap
superconductivity or superfluidity would appear in a temperature
window. Later, a more detailed discussion of the free energy was 
given,\cite{Heid} and it was shown that the different sign in the
gap equation also leads to the Meissner kernel having the wrong
sign.\cite{Colemanetal} This led to the abandonment of Berezinskii-type
odd-gap proposals\cite{Abrahamsetal} in the context of high-T$_{\rm c}$
superconductivity. Instead, these authors then considered more complicated
scenarios of composite order parameters that do not suffer 
from these problems.

In the present theory, these problems are solved by realizing that there
is a choice with respect to the reality properties of the gap function
we call $\Lambda_n$, and that the physical choice is to make it imaginary.
Let us recall the
observations that led to this realization. First of all, with
the conventional choice of a real $\Lambda_n$, the field theory that
results from expanding about a Fermi liquid saddle point is
unstable in the
particle-particle spin-triplet channel, which forces a deformation of
the integration contour. The validity of this procedure was then
checked by comparing the perturbation theory that results from the
field theory with ordinary perturbation theory within the framework
of a second quantized Hamiltonian. The comparison can be made in the
very simple case of noninteracting electrons, where there is no question
as to what the correct answer is. The perturbative result is also in
quantitative agreement with experiment. Next, in order for the saddle point
that describes the superconducting phase to lie on the deformed contour,
the saddle-point fields, which correspond to anomalous expectation values
in a fermionic formulation, must be chosen to be imaginary. This choice
turns out to minimize the saddle-point free energy, while the conventional
choice would maximize it. The conclusion
is that our interpretation of $\langle{^1_1Q}_{n-1,-n}\rangle$
as imaginary is correct, and there is no ambiguity or freedom left.
We note that the particle-particle spin-triplet channel is the only one
where the determination of the reality properties of the expectation values
is not straightforward.\cite{ModelFootnote} In all other channels, the
symmetry properties of the fluctuating fields, Eqs.\ (\ref{eqs:2.5}), agree
with the convergence requirements of the field theory, and no deformation
of the contour is necessary. We do not know why the particle-particle
spin-triplet channel is special in this respect.

One might object that the symmetry properties of
the $Q$ matrix, Eqs.\ (\ref{eqs:2.5}), just encode the properties of
the underlying fermionic fields, and must therefore also hold after taking
expectation values, which would make $\langle{^1_1Q}_{n-1,-n}\rangle$ real.
This objection is fallacious, however, due to the following reasons. The
symmetry properties hold for the fluctuating fields, which are just dummy
integration variables and have no direct physical meaning. Deforming the
integration contour to apply the method of steepest descent is therefore
permissible as long as one does not cross any singularities. This we have
checked be comparing with fermionic perturbation theory. The expectation
value $\langle{^1_1Q}_{n-1,-n}\rangle$, on the other hand, does have a
physical meaning, but it does not follow from Eqs.\ (\ref{eqs:2.5}) that
it must be real. In the absence of
spontaneous symmetry breaking, of course $\langle{^1_1Q}_{n-1,-n}\rangle = 0$. 
To obtain a non-zero expectation value, one
must consider a small field that is conjugate to ${^1_1Q}_{n-1,-n}$.
If this field were static, then hermiticity would indeed require
$\langle{^1_1Q}_{n-1,-n}\rangle$ to be real. However, in the particle-particle
spin-triplet channel the conjugate field cannot be static.
The reality properties of the
anomalous expectation value then depend on the properties of the unphysical,
time dependent, symmetry breaking field, which are not given a priori. They
could be determined from the requirement that 
$\langle{^1_1Q}_{n-1,-n}\rangle$ be imaginary, but we have not done that.
We also note that our saddle point violates only the hermiticity requirement,
Eq.\ (\ref{eq:2.5e}), but {\em not} the Pauli principle, Eq.\ (\ref{eq:2.5d}),
in agreement with the above discussion.

One might wonder why the static conductivity is affected by the existence
of a gap function, given that the latter is zero at zero frequency. The
answer lies in the fact that the infinite conductivity, like the Meissner
effect, is a direct consequence of the broken symmetry and the resulting
Goldstone mode.\cite{Weinberg} Indeed, in BCS theory with its constant
gap function, the prefactor of the delta-function in the conductivity is
independent of the gap. More specifically, given
the existence of a sound-like Goldstone mode that couples to the density
fluctuations, particle number conservation determines the form,
Eq.\ (\ref{eq:4.32b}), of the density susceptibility, which in turn
guarantees an infinite static conductivity. On an even more technical
level, we observe that while the Goldstone susceptibility, Eq. (\ref{eq:4.12}),
indeed vanishes at zero external frequency, the actual coupling to the
density fluctuations is more complicated. From Eq.\ (\ref{eq:4.31a}) we
see that the coupling is given by the
more general susceptibility $\chi^{(a)}_{12,34}$, Eq.\ (\ref{eq:4.13}),
multiplied by the eigenfunction $g$, which is a generalized gap function.
The net result, as far as frequency dependences is concerned, is the same
as for conventional superconductors.

We have neglected the explicit disorder term in the action, and have
taken the quenched disorder into account only implicitly, by using
the particle-particle spin-triplet interaction from Ref.\ \onlinecite{us_tsc}.
At the level of the Gaussian theory,
keeping the disorder explicitly just complicates the
calculations without leading to any essential changes, as we have pointed
out throughout the paper. For instance, the gap equation
is independent of the disorder (this is the spin-triplet
analog of Anderson's theorem), and the critical behavior is qualitatively
unchanged. Another (minor) modification is that the $\delta$-function
contribution to the conductivity no longer exhausts the f-sum rule, but
rather sits on top of a continuous spectrum. This conclusion from the
Gaussian theory would indeed be
correct if we could work in dimensions $D>2$. However,
Eqs.\ (\ref{eqs:2.7}) are valid only in $D=2$, and in $D>2$ the analogous
interaction does not lead to a superconducting ground state.\cite{us_tsc}
In $D=2$, on the other hand, the explicit disorder term in the action has
more subtle, qualitative consequences, which we discuss next.

\subsection{Outlook}
\label{subsec:V.C}

Throughout this paper, we have ignored the well-known fact that, in a
$2$-$D$ disordered interacting electron system at zero temperature, 
the paramagnetic normal metal phase is also unstable against the 
formation of an insulator, and a ferromagnetic phase. 
As a result, as we have stressed 
elsewhere,\cite{R} the nature of the ground state in the absence of
spin-flip scattering processes is not known, even in the absence of
the attractive spin-triplet interaction discussed above. 

We conclude that in a $2$-$D$ disordered electron fluid, there are at
least three competing instabilities of the normal metal ground state.
Which one of these wins is a priori unclear, and answering this question
would require a detailed and consistent study of the free energies for
the various phases, which has not been attempted so far.
It is certainly conceivable that different
instabilities dominate in different parameter regimes, leading to
very different ground states. It is also possible that the tendencies
towards forming an insulator and a spin-triplet even-parity
superconductor, respectively, effectively cancel one another, leading
to a normal metal-like phase at least in a certain temperature window.
This last possibility is the basic idea behind recent proposals to
explain the observed metal-insulator transition in Si MOSFETs and
other $2$-$D$ electron systems in terms of our exotic 
superconductivity.\cite{us_mosfets}
However, a better understanding of these issues will require much more
work.

\acknowledgments
This work was supported by the NSF under Grant Nos. DMR--95--10185,
DMR--96--32978, and DMR--98--70597. We gratefully acknowledge helpful
discussions with E. Abrahams, A. Balatsky, and D. Scalapino, and we 
especially thank
E. Abrahams for reading a preliminary version of the manuscript. This
work was completed during a stay at the Institute for Theoretical Physics
at the University of California at Santa Barbara, and was supported in part
by the National Science Foundation under grant No. PHY94--07194.

\appendix
\section{Weak-localization corrections, and the stability of the
 field theory}
\label{app:A}

The purpose of this appendix is to check that our field theory, defined by
the action and the integration contour specified in Sec.\ \ref{subsec:II.B},
correctly reproduces the results of standard perturbation theory. For this
purpose, we choose to calculate the well-known weak-localization corrections 
to the conductivity within our formalism.

To this end, we restore the disorder part of the
action, ${\cal A}_{\rm dis}$, although we will not need it explicitly.
It is well known that non-interacting disordered electrons in $D=2$ dimensions 
in the absence of spin-orbit scattering do not
have a metallic phase.\cite{LeeRama} In perturbation theory, starting from 
the Boltzmann value for the conductivity $\sigma$ and performing a disorder 
expansion, this phenomenon manifests itself as corrections at first order
that diverge logarithmically as the frequency $\Omega$
approaches zero. These corrections are of the form
\begin{equation}
\delta\sigma = \frac{1}{\epsilon_{\rm F}\tau}\,(a_{\rm s} + a_{\rm t})\,
               \ln \Omega\tau\quad.
\label{eq:A.1}
\end{equation}
Here $\epsilon_{\rm F}$ is the Fermi energy, $\tau$ is the elastic
scattering mean-free time, and the coefficients $a_{\rm s}$ and
$a_{\rm t}$ are due to contributions from the particle-particle spin-singlet
and spin-triplet channels, respectively. We need only the ratio of these
two coefficients. In the absence of any spin-flip
scattering processes as well as of magnetic fields, fermionic perturbation
theory yields\cite{AAKL}
\begin{equation}
a_{\rm t} = -3\,a_{\rm s}\quad.
\label{eq:A.2}
\end{equation}
This result has been confirmed by experiments on thin metallic 
films.\cite{Bergmann} By deliberately contaminating the sample with a
strong spin-orbit scatterer, the triplet contribution to the weak-localization
effect can be eliminated, which leads to a sign change of the effect as
well as a reduction of the amplitude by a factor of $1/2$. Eq.\ (\ref{eq:A.2})
is thus very well confirmed.

Perturbation theory within the $Q$-field theory\cite{us_fermions} yields
structurally
\begin{eqnarray}
\delta\sigma &\propto& \frac{-1}{V}\sum_{\bf p} \left[\left\langle
     -{^0_1(}\delta Q)({\bf p})\,{^0_1(}\delta Q)(-{\bf p})\right\rangle\,
     \tr s_0s_0\,\tr \tau_1\tau_1\right.
\nonumber\\
&&\qquad\left.+3\left\langle{^1_1(}\delta Q)({\bf p})\,{^1_1(}\delta Q)
                                                          (-{\bf p})
   \right\rangle\,\tr s_1s_1\,\tr \tau_1\tau_1\right]
\nonumber\\
&\propto& \frac{1}{V}\sum_{\bf p} \left[\left\langle
     {^0_1(}\delta Q)({\bf p})\,{^0_1(}\delta Q)(-{\bf p})\right\rangle\right.
\nonumber\\
&&\qquad\qquad\left.-3\left\langle{^1_1(}\delta Q)({\bf p})\,{^1_1(}\delta Q)
      (-{\bf p})\right\rangle\right]\quad.
\label{eq:A.3}
\end{eqnarray}
For notational simplicity we have suppressed the frequency labels on the
$Q$ matrices, which are related to the external frequency $\Omega$.
In the presence of disorder, the
$\langle Q\,Q\rangle$ correlation functions are diffusive,\cite{us_fermions} 
and therefore the momentum integral yields a $\ln\Omega\tau$ in $D=2$.
Comparing Eqs.\ (\ref{eq:A.1}) and (\ref{eq:A.3}) we see that the
field theory yields the same result as fermionic perturbation theory,
provided that the two correlation functions have the same sign and
magnitude. This is indeed the case with the procedure explained in
Sec.\ \ref{subsec:II.B}, or, equivalently, if one ignores any
convergence questions and formally performs the Gaussian integrals
over $Q$, as was done in Ref.\ \onlinecite{us_fermions}. We conclude
that the field theory, with the particle-particle spin-triplet matrix
elements integrated over the contour chosen in Sec.\ \ref{subsec:II.B}, 
correctly reproduces
perturbation theory, which in turn is in quantitative agreement
with experiment.


\begin{references}
\b{Berezinskii} V.L. Berezinskii, Pis'ma Zh. Eksp. Teor. Fiz. {\bf 20},
 628 (1974) [JETP Lett. {\bf 20}, 287 (1974)].
\b{us_mosfets} D. Belitz and T.R. Kirkpatrick, Phys. Rev. B {\bf 58}, 8214 
 (1998); D. Belitz and T.R. Kirkpatrick, Ref.\ \onlinecite{us_tsc}.
\b{Abrahamsetal} A. Balatsky and E. Abrahams, Phys. Rev. B {\bf 45},
 13125 (1992); E. Abrahams, A. Balatsky, J.R. Schrieffer, and P.B. Allen,
 Phys. Rev. B {\bf 47}, 513 (1993); E. Abrahams, A. Balatsky, D.J. Scalapino,
 and J.R. Schrieffer, Phys. Rev. B {\bf 52}, 1271 (1995).
\b{us_tsc} T.R. Kirkpatrick and D. Belitz, Phys. Rev. Lett. {\bf 66},
 1533 (1991);
 D. Belitz and T.R. Kirkpatrick, Phys. Rev. B {\bf 46}, 8393 (1992).
\b{NegeleOrland} J. W. Negele and H. Orland, {\it Quantum Many-Particle
 Systems} (Addison-Wesley, New York 1988).
\b{Notation_Footnote} We use the notation $a\cong b$ for "$a$ is isomorphic to
 $b$", $a\propto b$ for ``$a$ is proportional
 to $b$'', and $a\sim b$ for ``$a$ scales like $b$''.
\b{us_fermions} D. Belitz and T.R. Kirkpatrick, Phys. Rev. B {\bf 56}, 6513
 (1997); D. Belitz, F. Evers, and T.R. Kirkpatrick, Phys. Rev. B {\bf 58},
 9710 (1998).
\b{OP_Footnote} In Sec.\ \ref{sec:IV}, we will use $^1_1Q$ as the
 superconducting order parameter. If desirable, a different linear
 combination of $\psi\psi$ and ${\bar\psi}{\bar\psi}$, e.g. just a $\psi\psi$
 or a ${\bar\psi}{\bar\psi}$ alone, could be chosen as the order parameter.
 This would simply amount to a rotation in the space of bilinear products
 of fermion fields. An order parameter with a structure similar to $^1_1Q$
 appears in the Nambu formulation of BCS theory, see, e.g., J.R. Schrieffer,
 {\it Theory of Superconductivity}, Benjamin (Reading, Mass. 1983).
\b{PotentialFootnote} In order to keep our discussion technically as simple
 as possible, we consider only a special case of the spin-triplet attraction
 derived in Ref.\ \onlinecite{us_tsc}. Eqs.\ (\ref{eqs:2.7}) represent the
 potential resulting from the spin-fluctuation mechanism discussed in that
 reference, in the limit of a large spin susceptibility.
\b{FisherMaNickel} M.~E. Fisher, S.-K. Ma, and B.~G. Nickel, Phys.
 Rev. Lett. {\bf 29}, 917 (1972).
\b{Imaginary_Footnote} For general matrix elements, it is in a nontrivial
 direction in the complex plane. Concentrating on the anti-diagonal elements 
 ($n=-m-1$) just illustrates the point we want to make particularly well.
\b{DenneryKrzywicki} See, e.g., P. Dennery and A. Krzywicki, {\it Mathematics
 for Physicists} (Dover, Mineola N.Y. 1996), ch.4 Sec.16.
\b{Frequency_Footnote} The frequency constraint in the Kronecker deltas in
 Eqs.\ (\ref{eqs:4.1}) should actually read $\delta_{n_1,-n_2-1}$. Since the
 $-1$ is not important for any of our purposes, and disappears upon
 analytic continuation to real frequencies at zero temperature, we omit
 it here and throughout the remainder of Sec.\ \ref{sec:IV}.
\b{Gauge_Footnote} Note that this is not just a global gauge transformation,
 and therefore affects the physics.
\b{tsc_Footnote} Some of the intermediate equations in
 Ref.\ \onlinecite{us_tsc} have a factor of $i$ missing, but the final gap
 equation is correct.
\b{Heid} R. Heid, Z. Phys. {\bf 99}, 15 (1996).
\b{AndersonNambu} P.W. Anderson Phys. Rev. {\bf 112}, 1900 (1958);
 Y. Nambu, Phys. Rev. {\bf 117}, 648 (1960).
\b{ConservingApproximation_Footnote} This is useful because, for reasons that
 are not entirely obvious, our Gaussian approximation turns out to be
 conserving in the $r=3$ channel, but not in the $r=0$ one.
\b{ABModeFootnote} Since there is some confusion concerning this point in
 the literature, it is worthwhile pointing out that the Anderson-Bogolubov
 mode, as the present derivation makes obvious, is {\em not} the
 Goldstone mode, the latter being an unobservable mode in the particle-particle
 channel. Rather, it is a particle-hole channel manifestation
 of the Goldstone mode that comes about due to a coupling between the two
 channels.
\b{Colemanetal} P. Coleman, E. Miranda, and A. Tsvelik, Phys. Rev. B {\bf 49},
 8955 (1994).
\b{ModelFootnote} This statement is true for our model which considers a local
 field $Q({\bf x})$ only. If one allowed for nonlocal fields, then similar
 problems might arise in superconducting states with different order
 parameters.
\b{Weinberg} S. Weinberg, {\it The Quantum Theory of Fields}, vol.2
 (Cambridge University Press Cambridge 1995), ch. 21.6.
\b{R} For a review, see, D. Belitz and T.~R. Kirkpatrick, Rev. Mod. Phys.
 {\bf 66}, 261 (1994), Sec. X.B.5.
\b{LeeRama} For a review, see, e.g., P.A. Lee and T.V. Ramakrishnan,
 Rev. Mod. Phys. {\bf 57}, 287 (1985).
\b{AAKL} B.L. Altshuler, A.G. Aronov, D.E. Khmelnitskii, and A.I. Larkin,
 in {\it Quantum Theory of Solids}, I.M. Lifshits (ed.) (MIR, Moscow 1982),
 p. 130.
\b{Bergmann} G. Bergmann, Phys. Rep. {\bf 101}, 1 (1984).
\end{references}
\end{document}